\documentclass[%
reprint,
 amsmath,amssymb,
 aps,
]{revtex4-2}
\usepackage{CJK}
\usepackage{graphicx}
\usepackage{caption}
\usepackage{dcolumn}% Align table columns on decimal point
\usepackage{comment}
\usepackage{xcolor}
\usepackage[colorinlistoftodos]{todonotes}
\usepackage{hyperref}% add hypertext capabilities
\usepackage[english]{babel}
\usepackage{bm}% bold math
\usepackage{amsmath}
\usepackage{amssymb}
\usepackage{comment}
\usepackage{subfigure}
\usepackage{epstopdf}
\usepackage{hyperref}% add hypertext capabilities
\usepackage{appendix}
\usepackage{float}

\usepackage{mathtools}
\DeclarePairedDelimiter\bra{\langle}{\rvert}
\DeclarePairedDelimiter\ket{\lvert}{\rangle}
\DeclarePairedDelimiterX\braket[2]{\langle}{\rangle}{#1 \delimsize\vert #2}

\DeclareUnicodeCharacter{03B2}{\ensuremath{\beta}}
\UseRawInputEncoding

\begin{document}

\preprint{APS/123-QED}

%\title{Photoassociation of an Ultracold Atom-Molecule System}
\title{Triatomic Photoassociation in an Ultracold Atom-Molecule Collision}

\author{Ahmed A. Elkamshishy}
 \email{aelkamsh@purdue.edu}
 \affiliation{
 Department of Physics and Astronomy, Purdue University, West Lafayette, Indiana 47907 USA
}

\author{Chris H. Greene}%
 \email{chgreene@purdue.edu}
\affiliation{%
 Department of Physics and Astronomy, Purdue University, West Lafayette, Indiana 47907 USA
}%
\affiliation{Purdue Quantum Science and Engineering Institute, Purdue University, West Lafayette, Indiana 47907 USA}

\begin{abstract}
Ultracold collisions of neutral atoms and molecules have been of great interest since experimental advances enabled the cooling and trapping of such species. This study is a theoretical investigation of a low energy collision between an alkali atom and a diatomic molecule, accompanied by absorption of a photon from an external electromagnetic field. The long-range interaction between the two species is treated, including the atomic spin-orbit interaction. The long-range potential energy curves for the triatomic complex are calculated in realistic detail, while the short-range behavior is mimicked by applying different boundary conditions at the van der Waals length\cite{Le_Roy}.
The photoassociation (PA) rate of an atom colliding with a dimer is calculated for different alkali atoms, namely Na and Cs. The model developed in this study is also tested against known results for the formation rate of the Cs$_3$ complex via PA, namely to compare with Ref.\cite{O.Diliue_PRL,PA_CS2_exp}, and the results are in generally good agreement. 
\end{abstract}

\maketitle
%{\bf Ahmed:  please change all element symbols like Cs to a normal roman text font, instead of writing them in equation mode. You've already done that correctly in the Abstract.}
\section{\label{sec:level1}Introduction}
In recent years, promising strides have been made in the field of ultracold atomic and molecular collisions \cite{Trap,Trap2,Ni_tweezer_array} with many applications, such as reaction dynamics\cite{cold_chem,cold_cs3,chem_review_1} and applications to quantum computations\cite{Trap_QC}. Ultracold reactions can provide a suitable environment for studying quantum effects at temperatures lower than $1 mK$. At such temperatures, the quantum nature of the system dominates the interaction with small energy scales, such as spin-orbit and hyperfine splittings, which can add extensive complexity to the theoretical description. Furthermore, there has been significant advancement in the techniques of optical trapping of single alkali metal atoms\cite{Ni_4}. Such techniques can be utilized to achieve a high degree of quantum control of single atoms in optical tweezers\cite{Ni_3}, which could be used to create single molecules by photoassociation\cite{Ni_2,PA_review}. 
Trapped atoms at low energies are relatively simple to study since the collision is dominated by the s-wave dynamics\cite{Bohn_Julienne}. A diatomic molecule in its ground state can be created through a collision between the trapped atoms\cite{Ni_2,Ni_3}. Such processes in the ultracold start with two atoms colliding through an s-wave, with an external electromagnetic field driving a transition to an excited p-wave state in one of the atoms, creating a  complex that then decays (often with high probability) to the ground vibrational state of the molecule by spontaneous emission\cite{Bohn_Julienne,PA_review,Pillet_1997}. In short, photoassociation (PA) is a laser assisted collision in which the initial state has two (or more) species with relative collision energy $E$ above their s-wave threshold.\cite{Pillet_1997}.

The system of interest is a trapped atom and a molecule in two overlapping tweezers where the long-range atom-molecule interaction is dominant\cite{Le_Roy}. The method developed is applied to model collision between Na-NaCs, and Cs-NaCs. The molecule starts in its rovibrational ground state $\ket{X^1\Sigma;N = 0}$ where $N$ is the rotational quantum number for the molecule NaCs. Such process can be viewed as:
$A(P_{3/2})$ + NaCs$(\ket{X^1\Sigma,\nu = 0; N = 0}) + \hbar \omega \xrightarrow{} (A-$NaCs$)^*$, where A is either Na or Cs.
The interaction between the two species at low energy is controlled by the long-range effects between the two systems. The long-range interaction can be written as a multipolar expansion where each term comes as a power law $R^{-n}$ in a perturbative treatment.
The molecule NaCs has a permanent dipole moment and both systems are neutral. In Sec.\ref{sec:level2} the leading terms of the interaction Hamiltonian are investigated namely, the dipole-dipole interaction with a long-range behavior varying as $R^{-6}$, and the dipole-quadruple interaction with long-range behavior varying as $R^{-8}$. 

The main goal of the present study is to calculate the photoassociation rate of the triatomic system (atom + NaCs). In Sec.\ref{sec:level2} and \ref{sec:level3} a theoretical model is presented to treat the system. By choosing a convenient basis set, and using spectroscopic information about the dimer\cite{cs2_PECs}, approximate potential energy curves for the long-range atom-dimer interaction are calculated for different dissociative atomic channels ($S,P_{3/2}$), relevant to the initial and final PA states.
The transition rate is controlled by the electric dipole matrix element: $|\braket{i}{\vec{d}|f}|^2$. While complete knowledge of both the initial and final states would require the knowledge of the short-range behavior of the potential curves and eigenfunctions, which is inaccessible in our model treatment. However, since PA couples a low energy s-wave to a weakly bound p-state, the main contribution to the matrix element $\braket{i}{\vec{d}|f}$ comes from the long-range behavior. In this article, only long-range interactions are considered, and short-range effects are explored by varying the boundary conditions at some distance $R = R_0$ beyond which the long-range interaction is dominant. The values of the PA rate are given in Sec.\ref{subsec:level2}, and surveyed in \ref{sec:appendixB} for different values of the scattering length. As a check, the same model is applied to the Cs-Cs$_2$ system that was studied extensively in \cite{LepersI,LepersII,LepersIII}. Our Cs-Cs$_2$ rate calculations show general consistency with the experimentally recorded values\cite{O.Diliue_PRL,cold_cs3} for vibrational states at energies $E = (-0.762, -1.109) cm^{-1}$ relative to the atomic $P_{3/2}$ threshold. %{\bf Ahmed:  you need to explain the preceding sentence to me, and also give units to your energies here.}
%\textcolor{red}{In the preceding sentence, I am showing the consistency of our calculations by comparing with Jes{\'u}s and Dulieu's. The two numbers I mentioned are also mentioned in and Dulieu's PRL as a test against experimentally observed values which they also mentioned in their Fig2.b}

\section{\label{sec:level2}Theory and calculation methods}
\subsection{\label{sec:level}Interaction Hamiltonian}

\begin{figure}[h!]
\begin{frame}{}
  \includegraphics[width=1.0\linewidth]{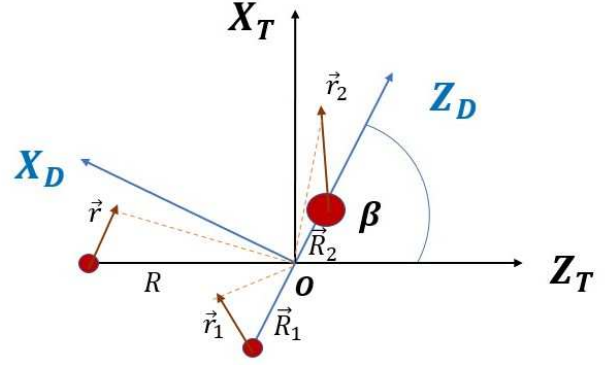} 
  \end{frame}
\caption{\label{fig:geometry} The blue coordinate system $\{X_D,Y_D,Z_D\}$ is chosen such that the dimer lies on the $Z_D$ axis. The black coordinate system   $\{X_T,Y_T,Z_T\}$ is chosen such that the axis connecting the center of mass of the atom and the dimer is the $Z_T$ axis. The two coordinates are related by a proper Euler rotation $R(0,\beta,0)$ around their mutual y axis with the angle $\beta$. In this convention, the y axis for both coordinates points out of the page.}
\end{figure}

The electrostatic energy between two charge configurations (A and B) can be written generally as (see Appendix.\ref{sec:appendixA} for details):
\begin{multline}
\label{int_pot}
    \hat{V}_{AB} =  \sum_{L_{A},L_{B}=0}^{\infty}\sum_{M=-L_{<}}^{L_{<}} f_{L_A,L_B,M} \frac{Q_{L_A}^M(\hat{r}_A) Q_{L_B}^{-M}(\hat{r}_B)}{R^{L_A +L_B +1}}
\end{multline}
Here, $R$ is the distance between the centers of mass of the two systems $A$ and $B$, as illustrated in Fig.\ref{fig:geometry}, $L_< =$ min$(L_A,L_B)$, and the operator $Q^L_M$ is the spherical multipole moment\cite{varsh}. The multipole moment for each system is expressed in its body-fixed frame, with its origin at the center of mass, as follows: 
\begin{equation}\label{mult_momement}
    Q_{L_A}^M(\hat{r}_A) = \sqrt{\frac{4 \pi}{2L+1}} \sum_{i\in A} q_i \hat{r}_i^{L_{A}} Y_{L_A}^{M}(\theta_i,\phi_i)
\end{equation} 
and similarly for system $B$. And finally:
\begin{multline}
    \label{f_consts}
f_{L_A,L_B,M} = 
(-1)^{L_B}\binom{L_A+L_B}{L_B+M}^{1/2}\binom{L_A+L_B}{L_B-M}^{1/2}
\end{multline}
Where $\binom{a}{b}$ is the binomial symbol.
The form of the constants $f_{L_A,L_B,M}$ depends on the orientation of the $Z_T$ axis in Fig.\ref{fig:geometry}. The convention used in this article is that system $A$ is an alkali atom, while system $B$ is the diatomic molecule.

Since both the atom and the molecule are neutral systems, the leading terms in the interaction in Eq.\ref{int_pot} start from $L = 1$. This study considers the dipole-dipole and dipole-quadrupole interactions, where $(L_A,L_B) = (1,1)$, and $(L_A,L_B) = (2,1)$, respectively. 
The nature of the interaction comes from the induced dipoles of the two systems, as well as the permanent atomic excited-state quadrupole moment. At very large distance, the interaction decays, but when the atom and the molecule get closer, the interaction dominates. The interaction potential in Eq.\ref{int_pot} is valid between two charge distributions as long as the two systems do not overlap\cite{Le_Roy,LepersI} and provided higher multipole contributions are negligible. 
The basis set used to diagonalize the interaction in Eq.\ref{int_pot} is chosen as $\ket{\Psi^{k}} = \ket{\psi_{a}^{k}}\ket{\psi_d^{k}}$. where $\psi_a$ and $\psi_d$ refers to the atom and the dimer unperturbed states, respectively.

\subsection{\label{subsec:level1} Constructing a basis set}

Alkali atoms like Na or Cs, with strong spin-orbit coupling\cite{alkali_model_dispersion}, are presented in the coupled basis with quantum numbers $\ket{\psi_a^k} = \ket{n,l,j,mj}$, where $k$ is a collective index for all quantum numbers. For a one-electron atom, the sum in Eq.\ref{mult_momement} becomes a single term. The operator $Q^L_M$ is a tensor operator whose matrix element between two atomic states is calculated using the Wigner-Eckart theorem\cite{varsh}.
\begin{multline}\label{atm_matrx_element}
\begin{split}
\braket{n',l',j',m_j'}{Q_L^M|n,l,j,m_j} & =\\
 -e\sqrt{\frac{4\pi}{2L+1}}C_{j m_j,L M}^{j' m_j'}C_{l 0,L 0}^{l' 0}\\
\times (-1)^{j+L+l'+1/2}\\
\times \sqrt{(2j+1)(2l+1)} 
\times \begin{Bmatrix}
l & 1/2 & j \\
j' & L & l'
\end{Bmatrix}\\
\times \int_0^{\infty} dr u_{n l j}(r)r^Lu_{n'l'j'}(r)\\
\end{split}
\end{multline}
The $C_{l_1 m_1,l_2 m_2}^{l_3 m_3} = \braket{l_3 m_3}{l_1 m_1;l_2 m_2}$ is a Clebsch-Gordan coefficient, $\begin{Bmatrix}
j_1 & j_2 & j_3\\
j_4 & j_5 & j_6
\end{Bmatrix}$ is the Wigner 6j symbol, and the function $u_{nlj}(r)$ is the radial part of the atomic wave function multiplied by $r$. The angular part is the same for any one-electron atom, while the radial part is the only difference between different atoms. 
A model potential for the electron in the atom is taken from\cite{alkali_model_dispersion}, and used to calculate the eigenstates. The atomic radial wave functions are calculated variationally using an appropriate Sturmian basis set\cite{SturmianII,sturmian}. The relevant atomic channels in the present study are ${S_{1/2},P_{1/2,3/2},D_{3/2,5/2}}$ with different principal quantum numbers. 

\begin{figure}[htp!]
    % \centering
\subfigure[]{\includegraphics[width=1\columnwidth]{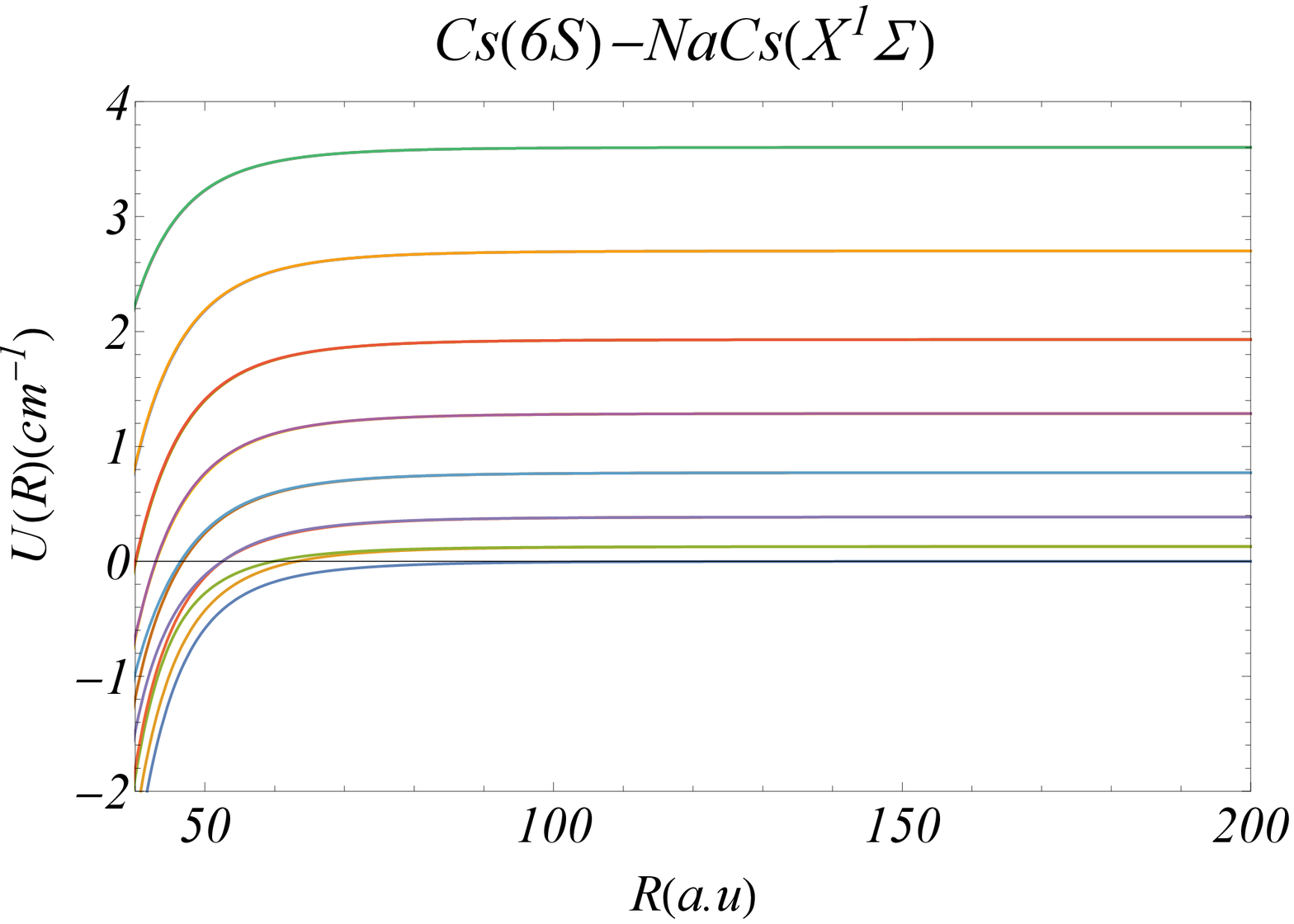}}
\subfigure[]{\includegraphics[width=1\columnwidth]{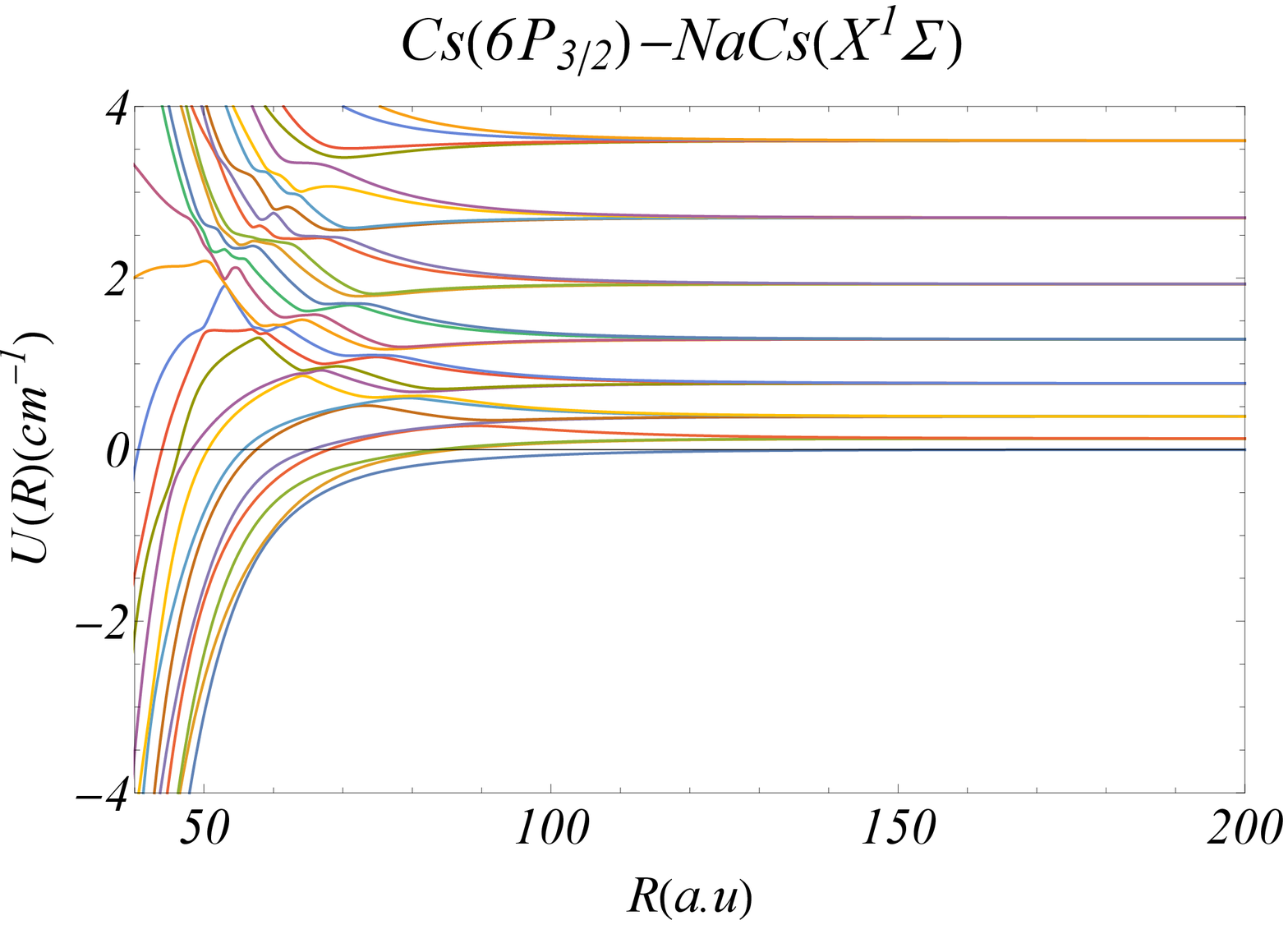}}
\subfigure[]{\includegraphics[width=1\columnwidth]{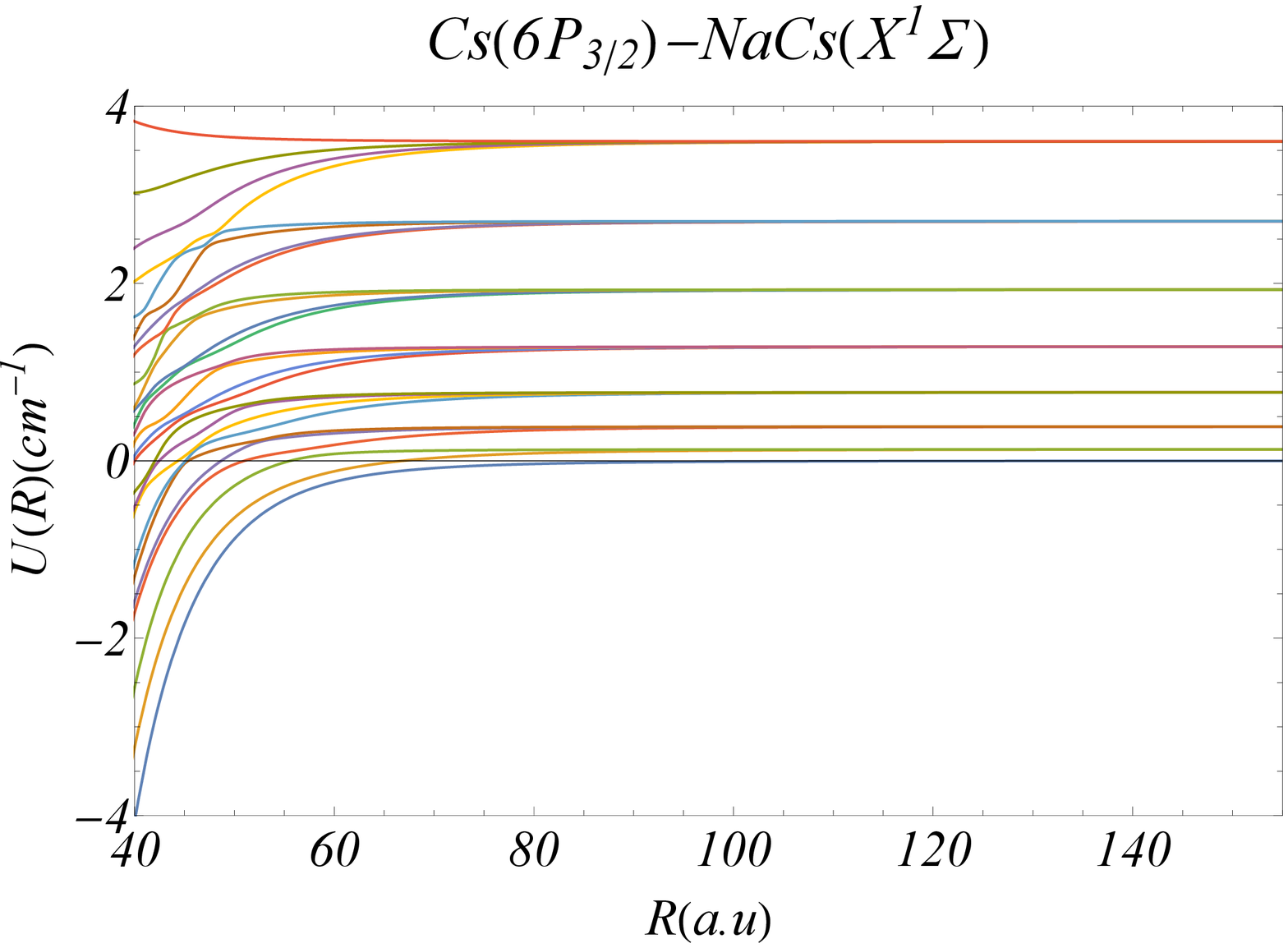}}
\caption{\label{fig:PEC cs}The potential energy curves for the Cs-NaCs system  are plotted as functions of the atom-dimer distance $R$. Each family of curves is associated with a single atomic channel, namely $\ket{6S;|\omega| =1/2}$ (a), and $\ket{6P_{3/2};|\omega| = 1/2}$ (b), and $\ket{6P_{3/2};|\omega| = 3/2}$(c). Each dissociative channel corresponds to a different dimer rotational quantum number $N$, in the range $N = 0-7$. The zero of the energy scale is fixed at the ground state energy for the independent atom-dimer system.}
\end{figure}
%%%%%%%%%%%%%%%%%%%%%%%%%%%%%%%%%%%%%%%%%%%%%%%%%%%%
\begin{figure}[tp]
    \centering
    \subfigure[]{\includegraphics[width=1\columnwidth]{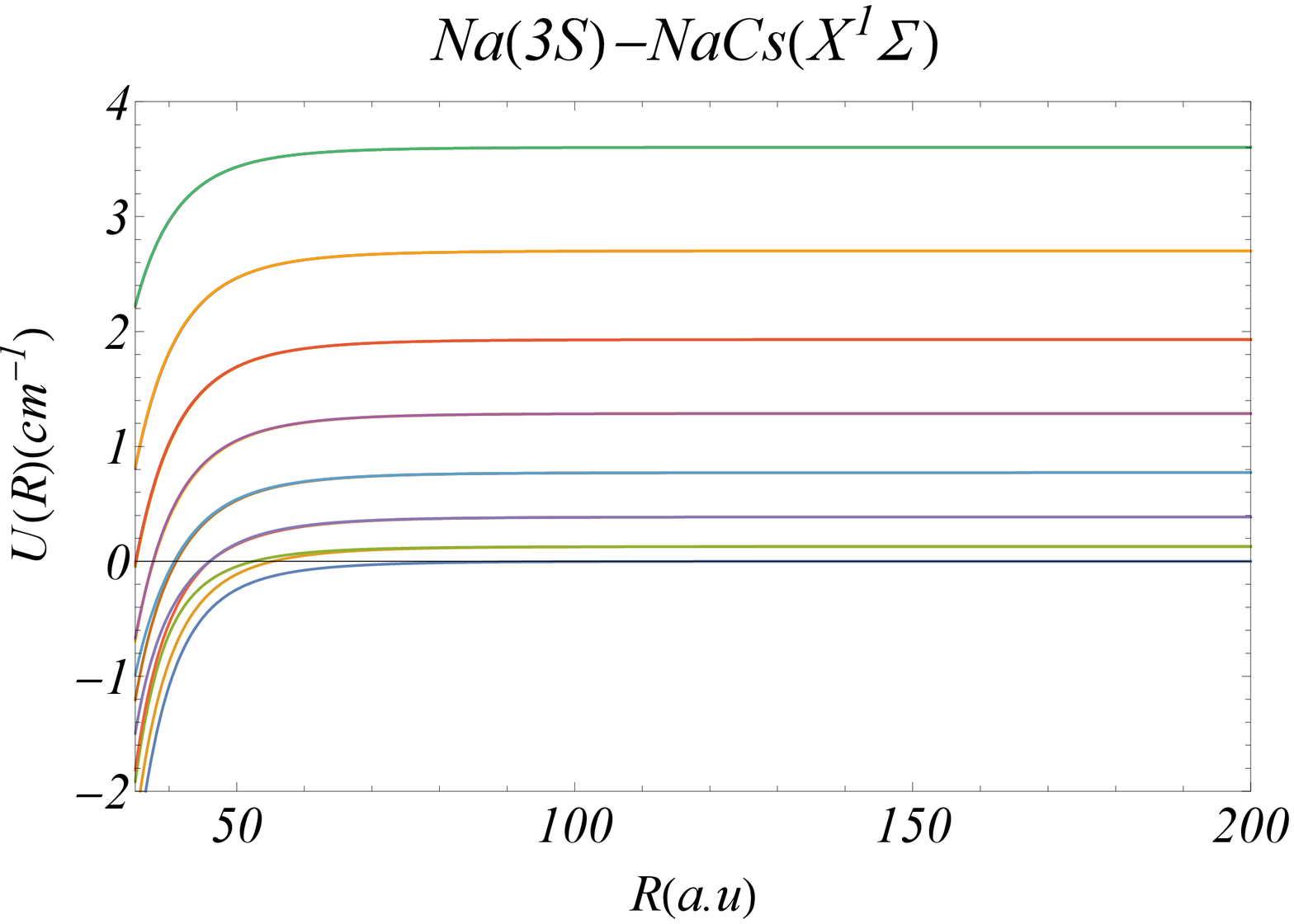}}
    \subfigure[]{\includegraphics[width=1\columnwidth]{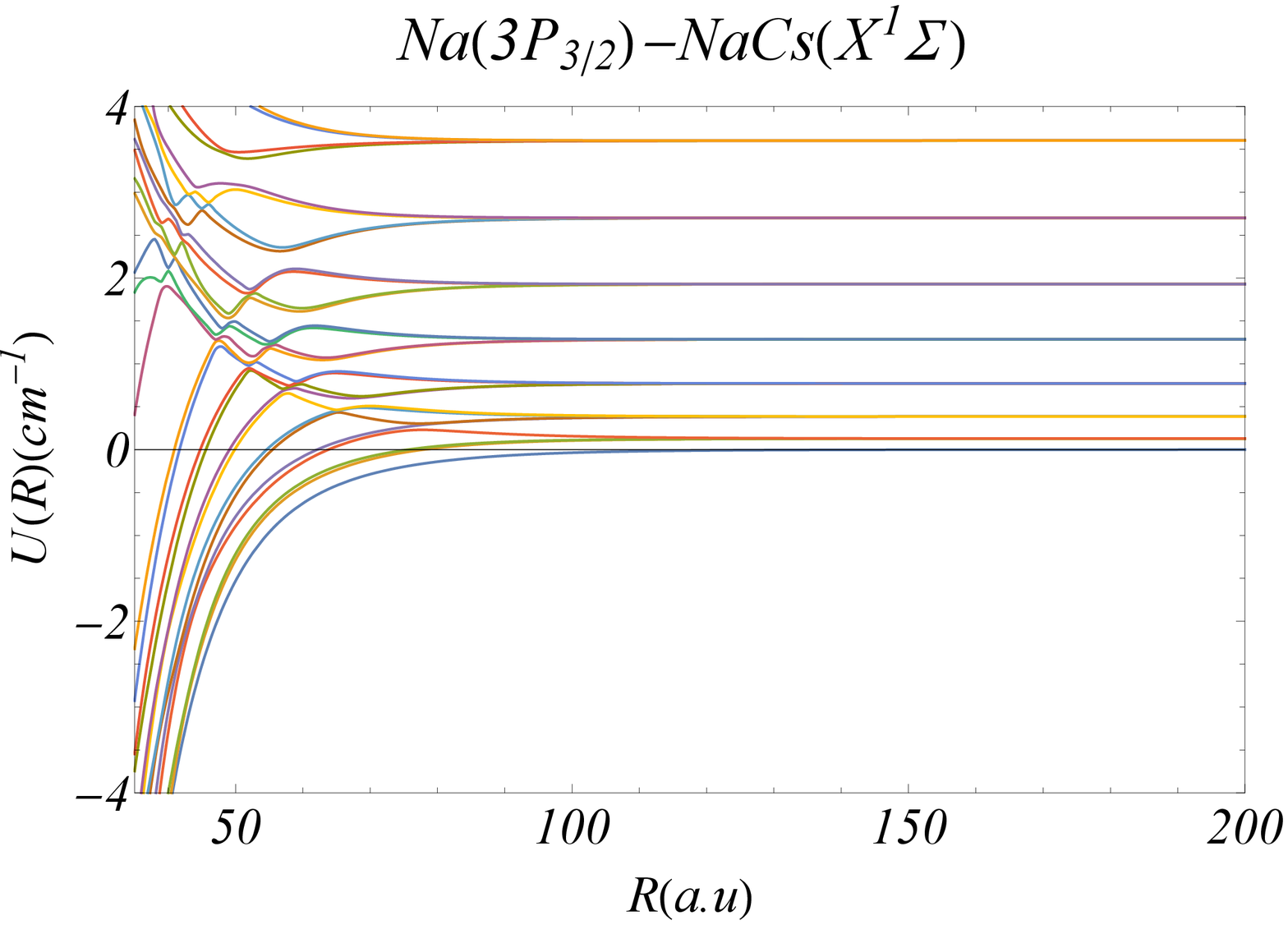}}
    \subfigure[]{\includegraphics[width=1\columnwidth]{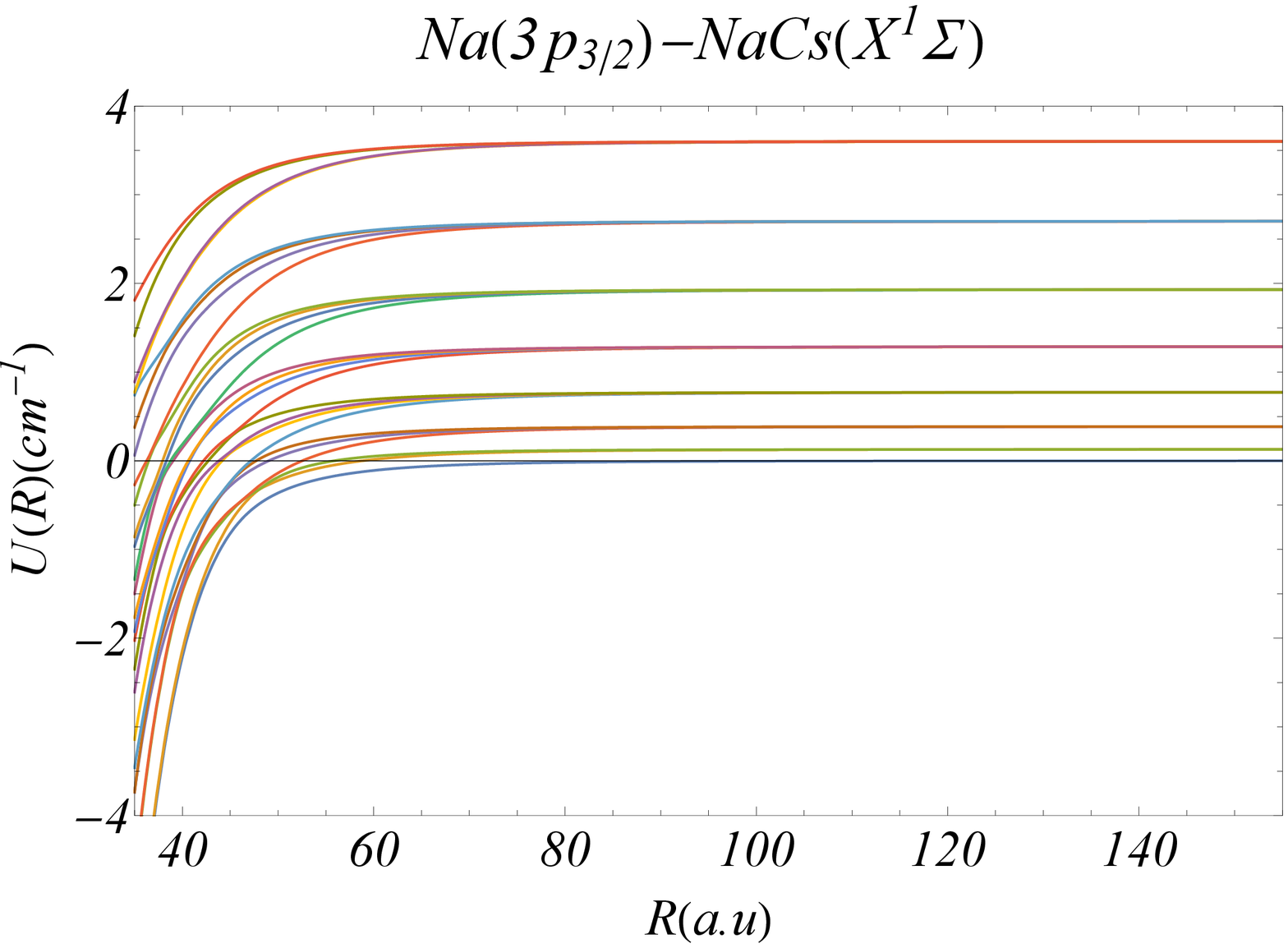}}
    \caption{The potential energy curves for the Na-NaCs system  are plotted as functions of the atom-dimer distance $R$. Each family of curves is associated with a single atomic channel, namely $\ket{3S;|\omega| =1/2}$ (a), and $\ket{3P_{3/2};|\omega| = 1/2}$ (b), and $\ket{3P_{3/2};|\omega| = 3/2}$(c). Each dissociative channel corresponds to a different dimer rotational quantum number $N$, in the range $N = 0-7$. The zero of the energy scale is fixed at the ground state energy for the independent atom-dimer system.}
    \label{fig:PEC na}
\end{figure}
%%%%%%%%%%%%%%%%%%%%%%%%%%%%%%%%%%%%%%%%%%%%%%%%%%%%%
The collision occurs between an atom and the ground state molecule. The basis set of the molecule is constructed using Hund's case C type states\cite{Hunds}, where the electronic spin-orbit interaction is strong. The total electronic angular momentum projection along the dime inter-nuclear axis ($Z_D$) $\Omega$ is added vectorially to the nuclear angular momentum projection. In the dimer body-fixed frame, with $Z_D$ defined as in Fig.\ref{fig:geometry}, the rotational wave function for a $\Sigma$ state, where $\Omega = 0$, is $Y_{N m}(0,\phi) = \sqrt{\frac{2N+1}{4 \pi}} \delta_{m,0}$. However, the two body-fixed frames are related by a single polar Euler angle of rotation $\beta$.
Transformation of rotational states between the diatom and the triatomic body-fixed frames is done using a lower-case Wigner rotation matrix $d(\beta)$\cite{varsh}, and the wave function in the trimer body-fixed frame takes the form $\psi_{N,m_N} = \sum_{m}d^N_{m m_N}\sqrt{\frac{2N+1}{4 \pi}} \delta_{m,0} =  \sqrt{\frac{2N+1}{4 \pi}}d^N_{m_{N},0}$. The symbol $d^N_{m,k}$ is an abbreviation for the matrix element of the rotation operator between two angular momentum states in the two related frames, namely $d^N_{m,k}(\beta) = \braket{N m}{d(\beta)|N k}$.
% The rotational states in both the diatom and trimer body-fixed frames are related  (Ahmed:  related to what?) by a a lower-case Wigner rotation matrix $d(\beta)$\cite{varsh},
In the triatomic body-fixed frame, the wave function of the dimer, in the Born-Oppenheimer approximation, is written as $\psi_{k}^{d} = \sqrt{\frac{2N+1}{4\pi}}d^N_{m_N \Omega}\Phi^{2e}_\Omega(\Vec{r}_1,\Vec{r}_2;\Vec{R}_{12}) \chi_\nu(R_{12})$ where $\Omega = 0,\pm{1}$ for $\Sigma$, and $\Pi$ states, respectively. $\Phi^{2e}_\Omega$ is the electronic wave function of the dimer, where $(\Vec{r}_{1},\Vec{r}_{2})$, and $\Vec{R}_{12} = \Vec{R}_1 - \Vec{R}_2$ are the coordinates for each electron, and the dimer inter-nuclear distance, respectively, as shown in Fig.\ref{fig:geometry}. Finally, the function $\chi_{\nu}(R_{12})$ is vibrational wave function that depends on the internuclear distance $R_{12}$. Potential energy curves for Na-Cs are taken from Ref.\cite{nacs_curves} for different $\Sigma$ and $\Pi$ symmetries that are relevant to electric dipole transitions.

The form of the interaction Hamiltonian in Eq.\ref{int_pot} uses the multiple moments $Q^L$ for both systems in their corresponding body-fixed frame. To evaluate matrix elements for the dimer, the operators need to be represented in the body frame of the triatomic system as well. Since $Q^L_M$ is a tensor operator\cite{varsh}, one can write 
\begin{equation} \label{rotation}
    (Q^L_M)^T = \sum_{M'}d^L_{M' M}(Q^L_M)^D
\end{equation}
where $(T,D)$ correspond to the two different frames illustrated in Fig.\ref{fig:geometry}. The multiple moment of the dimer in its body frame is given by Eq.\ref{mult_momement}, and finally the matrix element of the dimer multipole operator between two dimer basis states is given by
\begin{multline}\label{dimer_mtrx_element}
\braket{\psi^{k'}_d}{(Q^L_M)^T|\psi^k_d} = \sum_{M'}\braket{d^{N'}_{m_N'\Omega'}}{d^L_{M' M}|{d^N_{m_N \Omega}}} \\ \times \braket{\Phi^{2e}_{\Omega'}}{(Q^L_M)^D|\Phi^{2e}_\Omega}
\end{multline}
%%%%%%%%%%%%%%
%%%%%%%%%%%%%%%
The first term is evaluated using the properties of the Wigner matrix\cite{varsh} and is given by $\braket{d^{l_1}_{m_1 k_1}}{d^l_{m k}|{d^{l_2}_{m_2 k_2}}} = \sqrt{\frac{2l_2 +1}{2l_1 +1}}C_{l_2m_2,l m}^{l_1 m_1}C_{l_2 k_2, l k}^{l_1 k_1}$. An independent atom model is used to calculate the matrix element between two dimer states. The molecular states and energies are taken from the potential energy curves of Na-Cs in Ref. \cite{nacs_curves}. The basis set of the dimer has the dissociative atomic channels ${S+S,S+P,P+S}$, each with 8 rotational levels in the range $N = \Omega:\Omega+7$. The long-range Hamiltonian in Eq.\ref{int_pot} commutes with the total angular momentum projection of the system along the $Z_T$, thus $w = m_j + m_N$ is a good quantum number. The Hamiltonian matrix comes in a block-diagonal form with each block corresponding to a subspace of the unperturbed states with the same $w$.

\section{\label{sec:level3}Results}
\subsection{\label{subsec:level1} The long-range interaction energy}
The long-range potential energy curves are calculated by diagonalizing the interaction Hamiltonian in Eq.\ref{int_pot}, written in the body-fixed frame of the triatomic molecule for both Cs-NaCs Fig.\ref{fig:PEC cs} and Na-NaC Fig.\ref{fig:PEC na} systems, at different values of the atom-dimer distance $R$. The leading asymptotic term in the long-range dipole-dipole interaction gives the $C_6/R^6$ behaviour in the long-range potential energy curves.
The initial state $\ket{i}$ of PA has the quantum numbers  $\ket{S, X^{1}\Sigma;|\omega| = 0.5}$. Different set of final states are considered in this study, namely, $\ket{P_{3/2},X^{1}\Sigma; |\omega| = \frac{1}{2}}$ and $\ket{P_{3/2}, X^{1}\Sigma; |\omega| =  \frac{3}{2}}$. The radial part of the initial state $u_{E}(R)$ is a solution for $(\frac{-1}{2\mu}\frac{d^2}{dR^2} + V_S(R) - E)u_{E}(R) = 0$, with $E>0$, where $V_s(R)$ is the lowest curve for the $S$ family shown in Fig.\ref{fig:PEC cs}. The final vibrational states $f_{\nu}(R)$ are solutions of $(\frac{-1}{2\mu}\frac{d^2}{dR^2} +\frac{1}{\mu R^2} + V_{P_{3/2}}(R) - \epsilon_\nu)f_\nu(R) = 0$, with $\epsilon_\nu < 0$. The second term is the centrifugal barrier due to the total angular momentum of the final states $J = 1$. The index $\nu$ corresponds to different vibrational states.
%{\bf Ahmed:  please try to be more consistent in your notation.  Sometimes you use a lower case s,p for the atomic states, and other places you use capital S,P}

 \begin{figure}[b]
    \centering
    \captionsetup{width=1\columnwidth}
    \subfigure[]{\includegraphics[width=1\columnwidth]{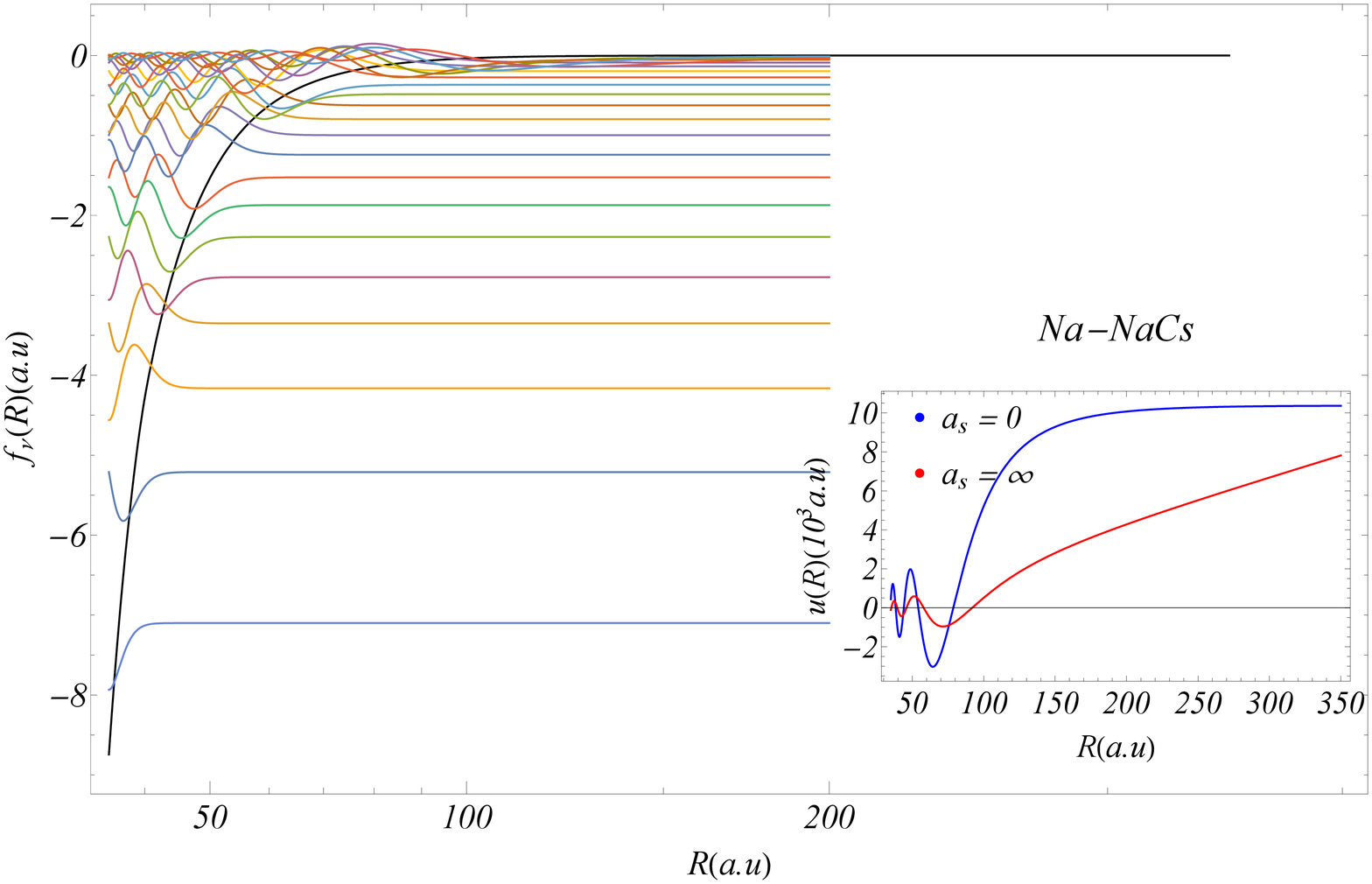}}
    \subfigure[]{\includegraphics[width=1\columnwidth]{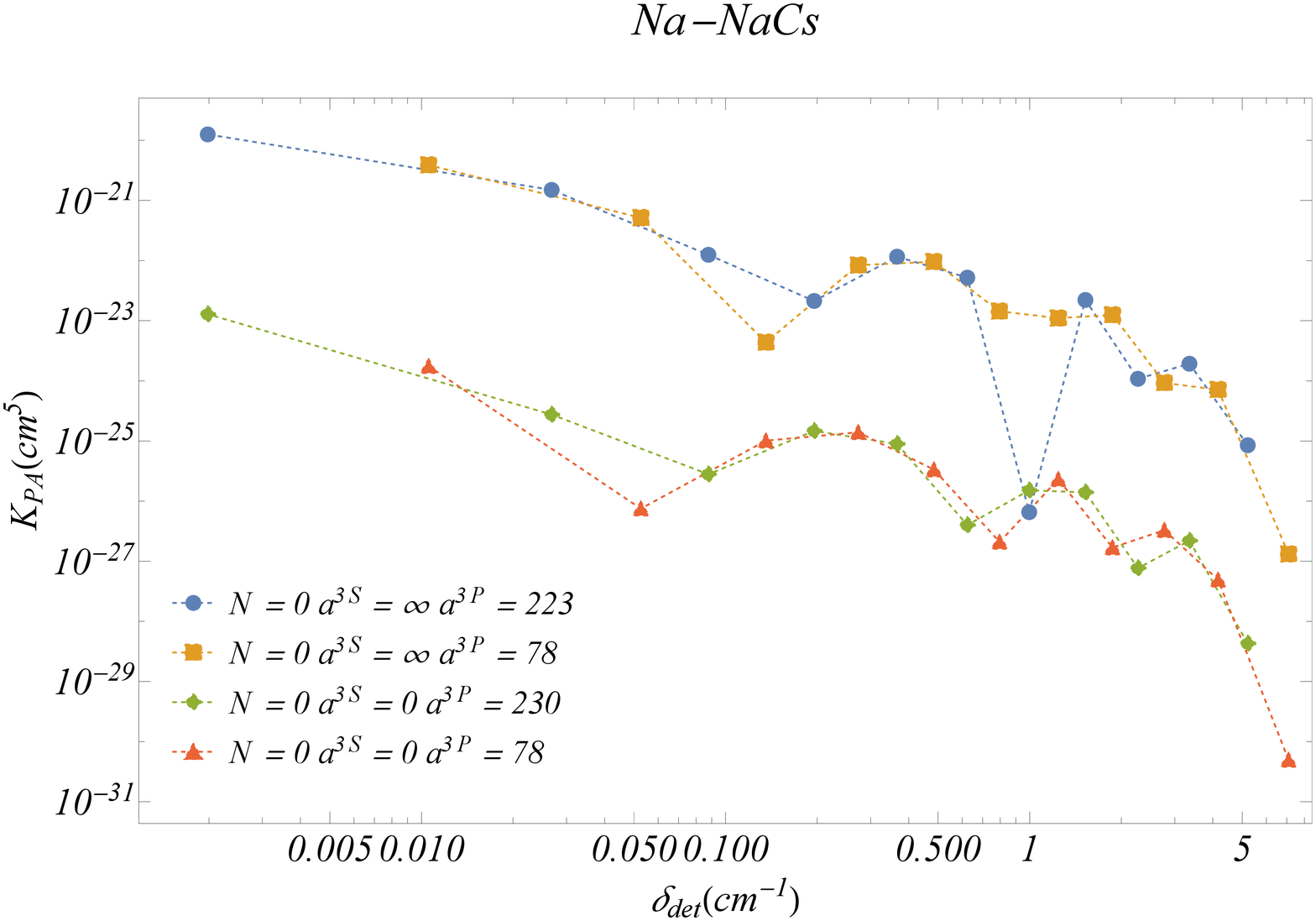}}
    \caption{(a) The spectrum and the wave functions of the final states $\ket{3P_{3/2};N = 0;\epsilon_{\nu}<0;|\omega| = 1/2}$ are plotted vs the internuclear distance $R$. Two different boundary conditions considered, namely $a = 223$, and $a = 78$. the inset shows the energy normalized radial wave functions for the initial state of Na-NaCs, $\ket{3S;N = 0;E}$. The blue curve is the state with infinite scattering length while the red curve is the state with zero scattering length. (b) The normalized PA rate $K_{PA}$ is plotted for each final state at average collision energy $T = 200 nK$. Each color corresponds to different values for the scattering lengths as shown in the inset.}
    \label{fig:PA_Na}
\end{figure}

\begin{figure}[htp!]
    \centering
    \captionsetup{width=1\columnwidth}
    \subfigure[]{\includegraphics[width=1\columnwidth]{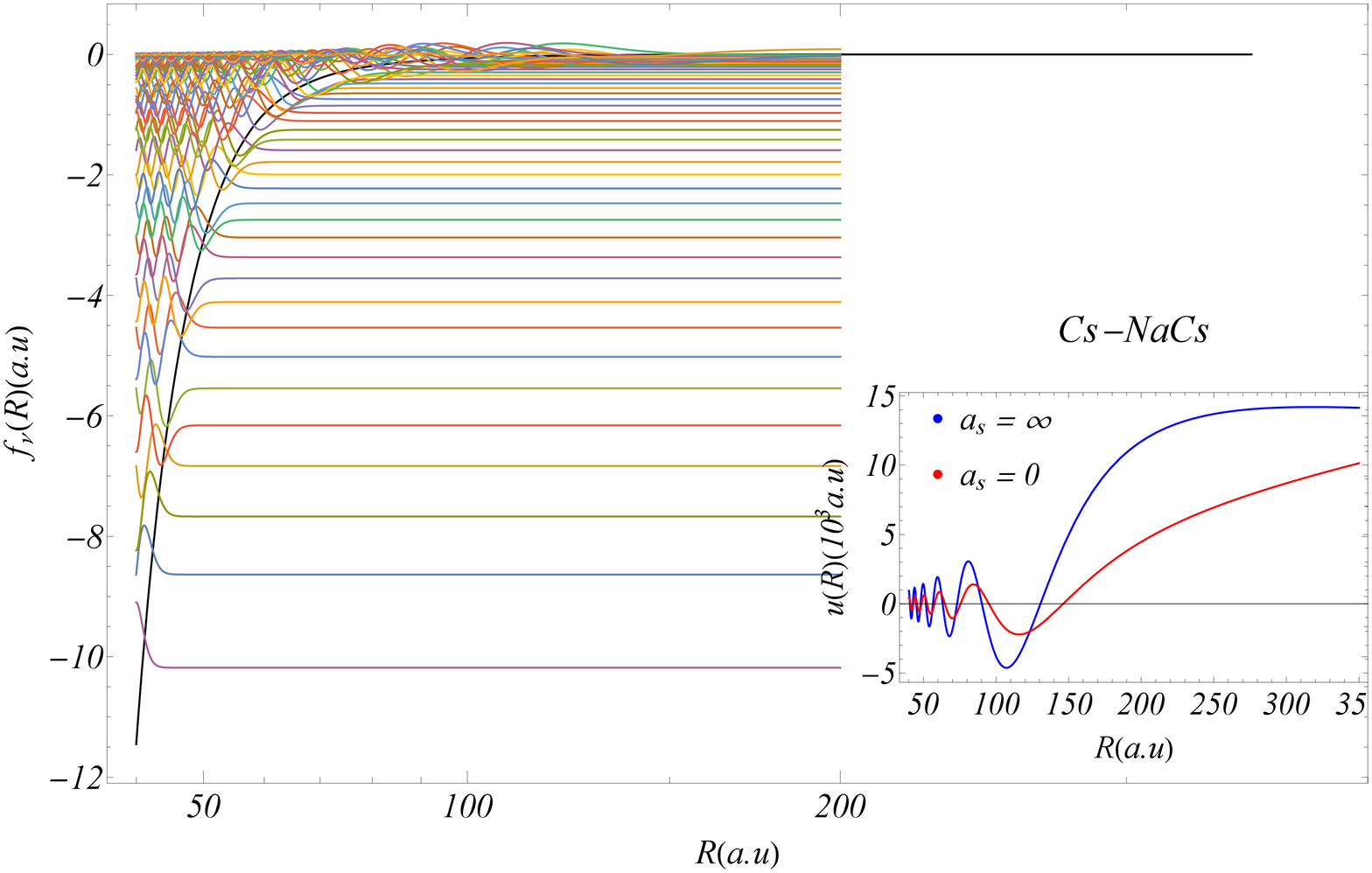}}
    \subfigure[]{\includegraphics[width=1\columnwidth]{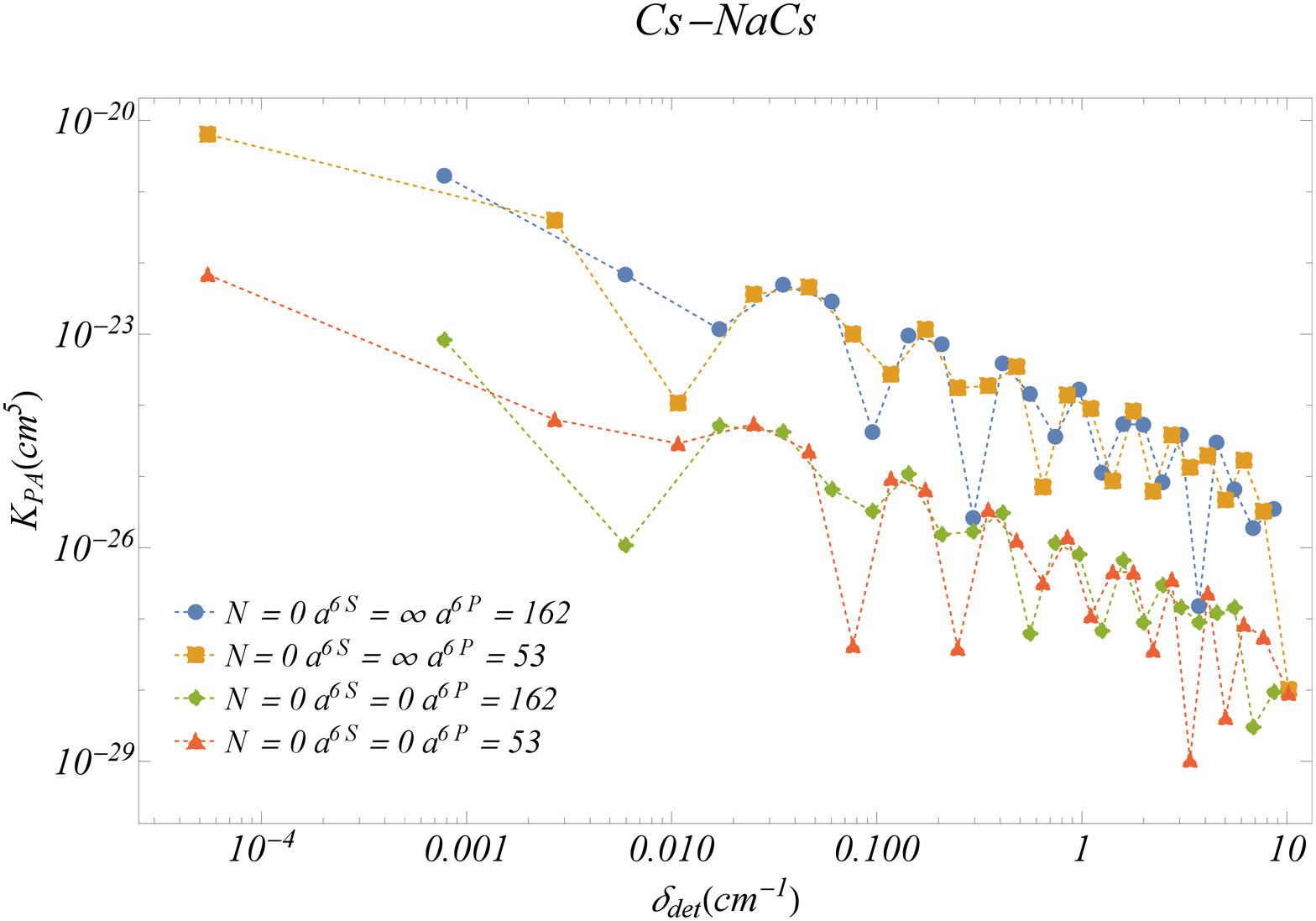}}
    \caption{(a) The spectrum and the wave functions of the final states $\ket{6P_{3/2};N = 0;\epsilon_{\nu}<0;|\omega| = 1/2}$ are plotted vs the internuclear distance $R$. Two different boundary conditions considered, namely $a = 162$, and $a = 53$. the inset shows the energy normalized radial wave functions for the initial state of Cs-NaCs, $\ket{6S;N = 0;E}$. The blue curve is the state with infinite scattering length while the red curve is the state with zero scattering length. (b) The normalized PA rate $K_{PA}$ is plotted for each final state at average collision energy $T = 200 nK$. Each color corresponds to different values for the scattering lengths as shown in the inset.}
    \label{fig:PA_CS}
\end{figure}

\subsection{\label{subsec:level2} Photoassociation rates}
For a system of interacting species with density $n_{mol}$, in thermal equilibrium at temperature T, the thermally averaged photoassociation rate $R_{PA}$ (the number of triatomic molecules formed per unit time) is given by\cite{Pillet_1997,O.Diliue_PRL}
\begin{multline} \label{PA_rate}
    R_{\nu}(E,T) = \frac{h^2}{2}n_{mol}\frac{1}{(2\pi\mu k_B T)^{3/2}}\frac{8\pi I_{PA}}{c} |S_\nu(E)|^2 e^{-\beta E}
\end{multline}
where $S_\nu(E) = \bra{\psi_i(E)}\vec{d}\cdot \hat{\varepsilon}\ket{\psi_f(\nu)}$, $\hat{\varepsilon}$ is the laser polarization, and $I_{PA}$ is the laser intensity. The energy $E = \epsilon_{\nu} - \hbar \omega_L$ is the energy of the initial continuum state, where $\epsilon_{\nu}$ is the energy of the final state and $\omega_L$ is the laser frequency. The factor $e^{-\beta E}$ selects only the energies within the same order of magnitude as $k_B T$, i.e. $e^{-E/k_B T} \xrightarrow{ E/{k_BT} >> 1}0$. The integral $S_\nu(E)$ is controlled by the value of the potential at the Franck-Condon(FC) point $R_C$ where only vertical transitions occur with high probability.\cite{Bohn_Julienne}. Although the expression of  $S_\nu(E)$ depends on the the initial collision energy, the initial continuum state does not change significantly around the FC points for different energies. This could be seen by looking at the energy normalized WKB wave function $\psi_{WKB} \approx \sqrt{\frac{2\mu}{\pi k(R)}} \sin(\int k(R)dR) + \phi$, where $k(R) = \sqrt{\frac{2\mu}{\hbar ^2}(E-V(R))}$. The potential energy, at distances up to the FC point, $\frac{V}{k_B}\approx 1 mK$ is much larger than the temperature of the system, and since PA is only significant for small $E$, the wave function becomes approximately energy independent. We fix the initial collision energy at $E = k_B T$ while calculating the initial wave function.
The expression in Eq.\ref{PA_rate} follows the perturbative treatment shown in detail in \cite{Pillet_1997} which requires calculating the radial wave function of the initial and the final states.
Alternatively, in the Bohn and Julienne treatment Ref\cite{Bohn_Julienne}, the matrix element $\bra{\psi_i(E)}\vec{d}\cdot \hat{\varepsilon}\ket{\psi_f(\nu)}$, characteristic of the line shape of PA processes, is given in terms of quantities that require only the calculation of the potential energy curves. Their treatment is a semi-analytical method that treats different laser processes including multi-photon processes using multi-channel quantum defect theory (MQDT). The radial wave function of the triatomic system depends on the short-range behavior of the potential curves which is not described by Eq.\ref{int_pot}. The short-range effects are taken into account to some degree by applying different boundary conditions at a distance $R = R_0$ beyond which the long-range interaction dominates. At low energy, the strength of the PA rate is controlled by the $S$-wave scattering length $a$ of the initial state for which the boundary conditions are tuned such that $a = 0$ and $a = \infty$ to see two limiting extreme values. On the other hand, the final states are weakly bound with turning points at long distance beyond $R = R_0$ where the main contribution comes from the long-range potential. The spectrum of the final states depends on the boundary conditions applied at $R_0$ for which two limits are also considered. The first boundary condition is tuned to give the minimum value of the lowest energy, while the second boundary condition gives the maximum value of the lowest energy; accordingly our calculations cover an upper and lower bound on the final state energies for any value of the scattering length. While PA can occur for different dimer collision channels, the matrix element $\bra{\psi_i(E)}\vec{d}\cdot\hat{\varepsilon}\ket{\psi_f(\nu)}$ has a non-zero value at large distance when the transition takes place between the same dimer state (in this case $N = 0$) which gives a strong PA rate as shown in Figs.\ref{fig:PA_Na}, \ref{fig:PA_CS}, and \ref{fig:PA_wf = 1.5} (See Appendix.\ref{sec:appendixB} for higher rotational levels).

 \begin{figure}[b]
    \centering
    \captionsetup{width=1\columnwidth}
    \subfigure[]{\includegraphics[width=1\columnwidth]{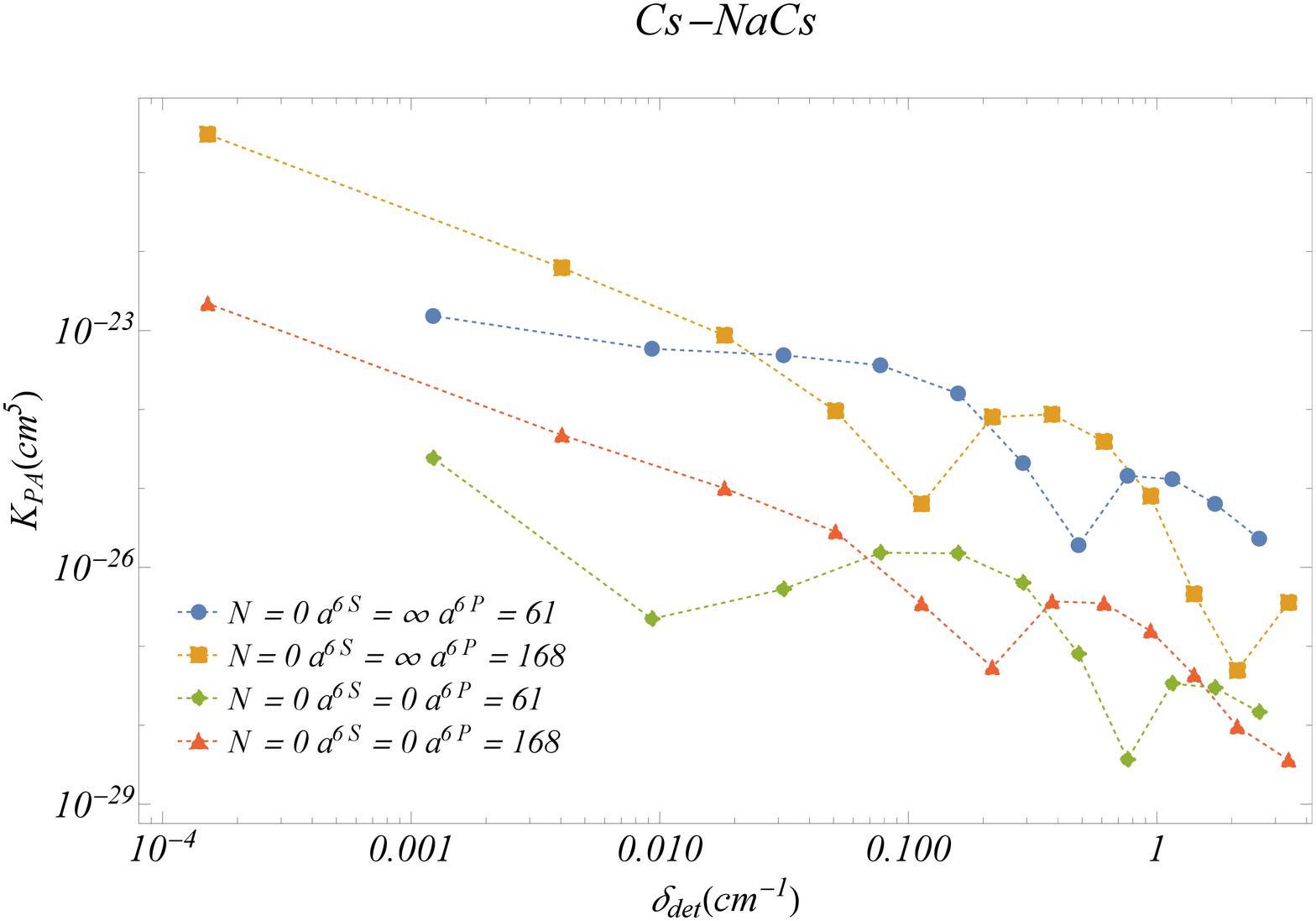}}
    \subfigure[]{\includegraphics[width=1\columnwidth]{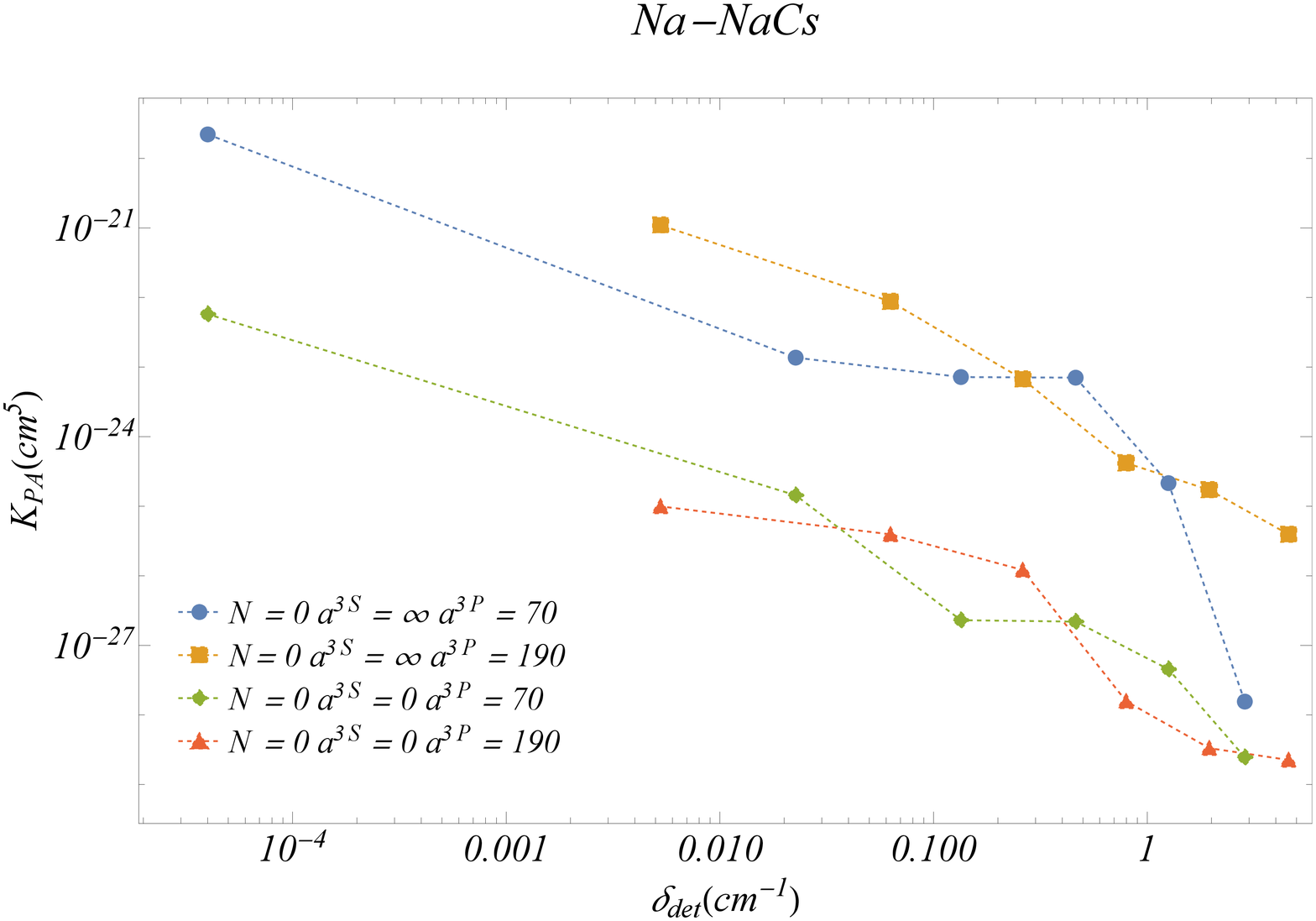}}
    \caption{The normalized PA rate $K_{PA}$ is plotted versus the vibrational energies of the final states $\ket{P_{3/2};N = 0;\epsilon_{\nu}<0;|\omega| = 3/2}$ for Cs-NaCs (a) and Na-NaCs (b) at average collision energy $T = 200 nK$. The initial state of PA is the same as the one in Figs.\ref{fig:PA_CS},\ref{fig:PA_Na},$\ket{S;N = 0;E}$ and similarly, the four colors correspond to different scattering lengths. }
    \label{fig:PA_wf = 1.5}
\end{figure}

The model developed in this study and described in Sec.\ref{sec:level1} ignores transitions between different vibrational states of the diatomic molecule. Moreover, the two-electron dimer wave function used in Eq.\ref{dimer_mtrx_element} is approximated by two independent atom wave functions. Such an approximation may not be accurate for calculations of different quantities such as the molecular polarizability. For alkali dimers like Cs$_2$, the vibrational transitions contribute significantly to the  ground state molecular polarizability \cite{cs3_pola}. More calculations of the dynamical polarizabilities that include vibrational transitions can be found in Ref.\cite{LepersII,LepersIII}nevertheless, the emphasis in this article is on the long-range behavior of the potential energy curves. The model developed in this study is used to calculate PA rate of Cs-Cs$_2$ system at $T = 500 nK$ as shown in Fig.\ref{fig:PA_Cs3}, and the results are in a good agreement with ones obtained in \cite{O.Diliue_PRL}. 

\section{Conclusion}
\begin{figure}[htp!]
    \centering
    {\includegraphics[width=1\columnwidth]{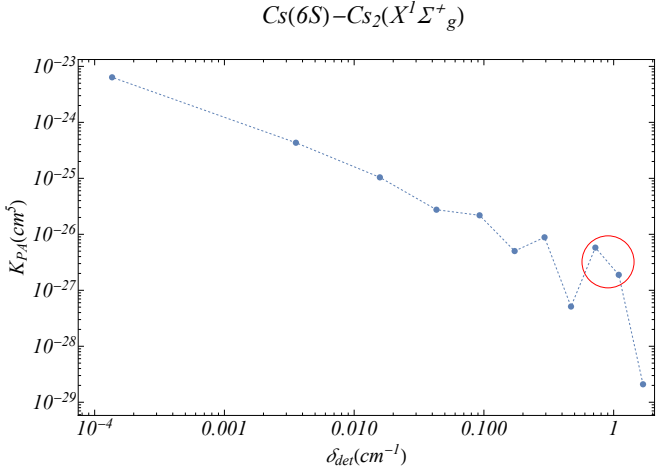}}
    \caption{ The PA rate of the Cs-Cs$_2$ system is calculated between the initial state $\ket{6S;N = 0;E}$ at $T = 500 nK$, and the final states with the symmetry $\ket{6P_{3/2};N = 0;\epsilon_n}$ for scattering length $a = 680$. The two points in the red circle agree with the ones observed experimentally at vibrational energies $E = -0.762 cm^{-1}$ and $E = -1.019 cm^{-1}$ and recorded at Ref.\cite{O.Diliue_PRL,cold_cs3}. Fig.2a in Ref.\cite{O.Diliue_PRL} has the PA rate plotted for final states with $N = 0,2$. Here, we focus on two points with experimental realization to test the validity of the treatment in this article.}
    \label{fig:PA_Cs3}
\end{figure}

This article has developed and applied a model to treat triatomic photoassociation of cold alkali atoms. Such process is governed by the long-range interaction among the colliding species, which is dominated by the electric dipole and quadruple interactions between the atom-dimer system. The model is used to calculate the long-range potential energy curves for our systems of interest (Na-NaCs and Cs-NaCs) and to calculate the PA rate of the triatomic system.  In the low energy regime, the resonance profile of these transitions is controlled primarily by the $S$-wave scattering length. In Sec.\ref{sec:level2}, we focus on atomic transitions to the states where the dimer rotational state is the same. In Appendix.\ref{sec:appendixB} transitions to different rotational levels are studied and shown in detail for both systems of interest, with emphasis on the dependence of the PA rate on the atom-dimer scattering length. 
Furthermore, the results presented in this article underline the difference among polar (non-polar) species, NaCs (Cs$_2$) in the ultracold PA process. The main difference is evident from comparing the photoassociation rates of the Cs-NaCs and Cs-Cs$_2$ systems, shown in Figs.\ref{fig:PA_CS} and \ref{fig:PA_Cs3}. For molecules with a strong permanent dipole moment, the density of the bound vibrational states for the triatomic system is higher than for trimers associated with non-polar dimers, and that allows for more transitions to a long-range vibrational state. In short, atom-molecule PA is enhanced for polar molecules, and the permanent dipole makes the long-range quadrupole-dipole interaction between the atom-dimer stronger and more dominant than the case of non-polar dimers.
\label{sec:Conclusion}

\section{Acknowledgment}
We thank Jes{\'u}s P{\'e}rez-R{\'i}os for informative discussions. This work is supported by the AFOSR-MURI, grant number FA9550-20-1-0323.

\appendix
\section{Derivation of the long-range potential}
\label{sec:appendixA}
In this section, we sketch a derivation of the expression for the long-range electrostatic interactions between two charge configurations. The first step writes the electrostatic energy between two discrete charge distributions A and B. Charges ($q_i$,$q_j$) in system ($A$,$B$) have coordinates ($\Vec{r}_i$,$\vec{r}_j$). The interaction energy is written in atomic units as:
\begin{equation}\label{energy_charges}
    U = \sum_{i\in A}^{} \sum_{j\in B}^{} \frac{q_iq_j}{r_{ij}},
\end{equation}
where $r_{ij} = |{\vec{r}_i - \vec{r}_j}|$. From Fig.\ref{fig:geometry}, $r_{ij}$ can be written as $r_{ij} = |\vec{r}_j+\vec{R}-\vec{r}_i|$. Using a spherical expansion of $r_{ij}^{-1}$\cite{harmonics}:
\begin{equation}\label{laplace_expansion}
    \frac{1}{|\vec{r}_j + \vec{R} - \vec{r}_i|} = \sum_{l,m} (-1)^m I_l^{-m}(\vec{r}_j+\vec{R}) R_l^m(\vec{r_i})
\end{equation}
where $(R_l^m,I_l^m)$ are the regular and the irregular harmonics given by $\sqrt{\frac{4\pi}{2l+1}}Y_{l}^{m}(r^l,\frac{1}{r^{l+1}})$. Expanding $I_l^m$ gives\cite{varsh}:
\begin{multline}\label{irregular_harmonics_expansion}
    I_l^m(\vec{r}_j+\vec{R}) = \sum_{l',m'} R_{l'}^{m'}(\vec{r}_j) I_{l+l'}^{m-m'}(\vec{R})\\
    \times{\binom{l + l'-m + m'}{l' + m'}}^{1/2} {\binom{l + l'+ m - m'}{l' - m'}}^{1/2}
\end{multline}
Now, choosing $\vec{R}$ to lie on the $Z$ axis, this implies $I_l^m(Z,0,\phi) = I_l^0(Z,0,0) = \frac{1}{Z^{l+1}}$. Combining the latter with both Eq.\ref{laplace_expansion}, \ref{irregular_harmonics_expansion} gives
\begin{multline}\label{spherical_expansion_electrostatic}
U = \sum_{l,l',m}\binom{l+l'}{l'+m}^{1/2}\binom{l+l'}{l'-m}^{1/2}\\
\times(-1)^{l'}\frac{\sum_{i\in A}^{}q_iR_l^m(\vec{r}_i) \sum_{j\in B}^{}q_j R_{l'}^{-m}(\vec{r}_j)}{R^{l+l'+1}}
\end{multline}
The sum over $(i,j)$ gives the spherical multipole moments of systems $(A,B)$ with an origin at each system's center of mass giving the same expression in Eq.\ref{int_pot}
 \begin{figure}[htp]
    \centering
    \captionsetup{width=1\columnwidth}
    \subfigure[]{\includegraphics[width=1\columnwidth]{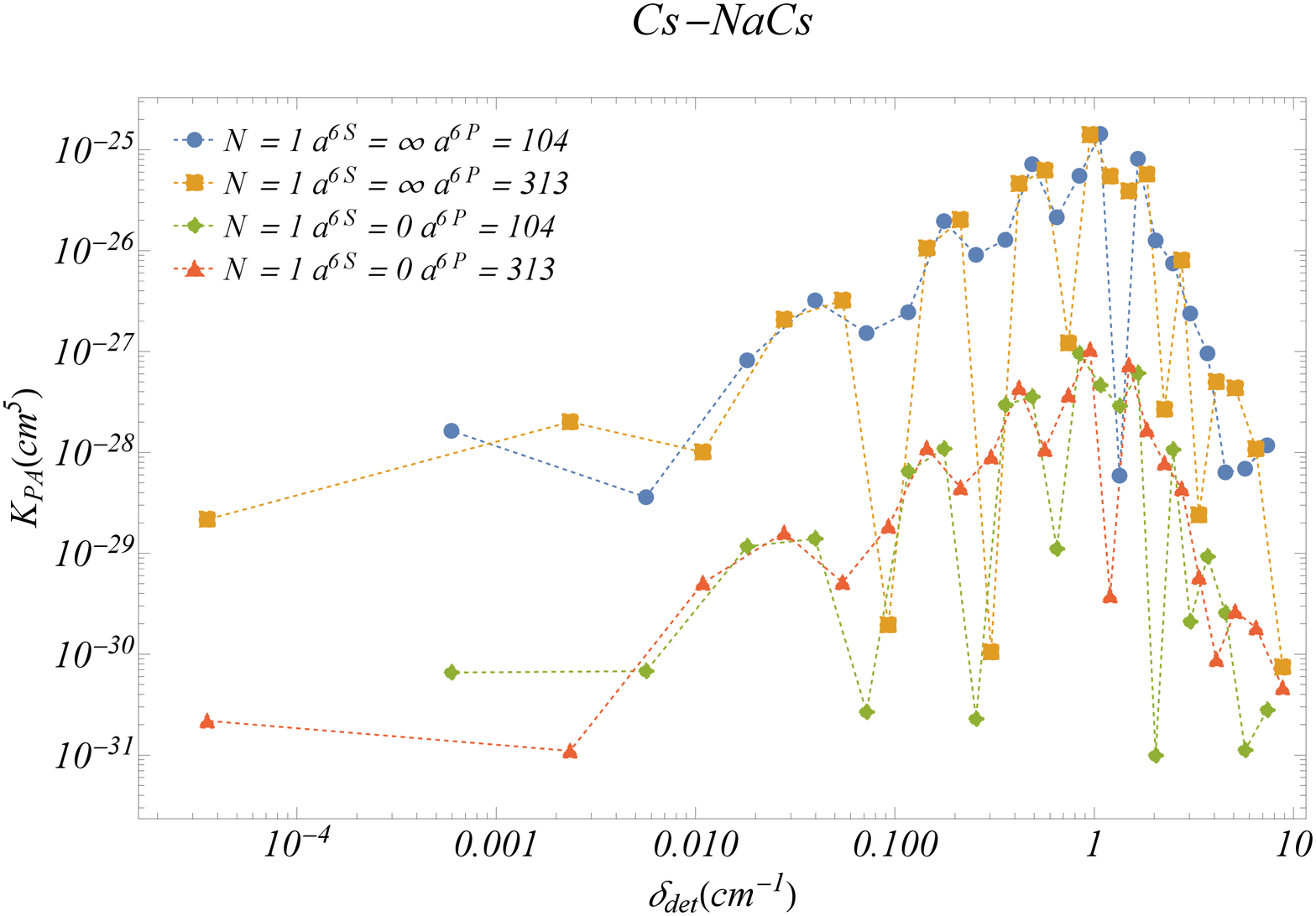}}
    \subfigure[]{\includegraphics[width=1\columnwidth]{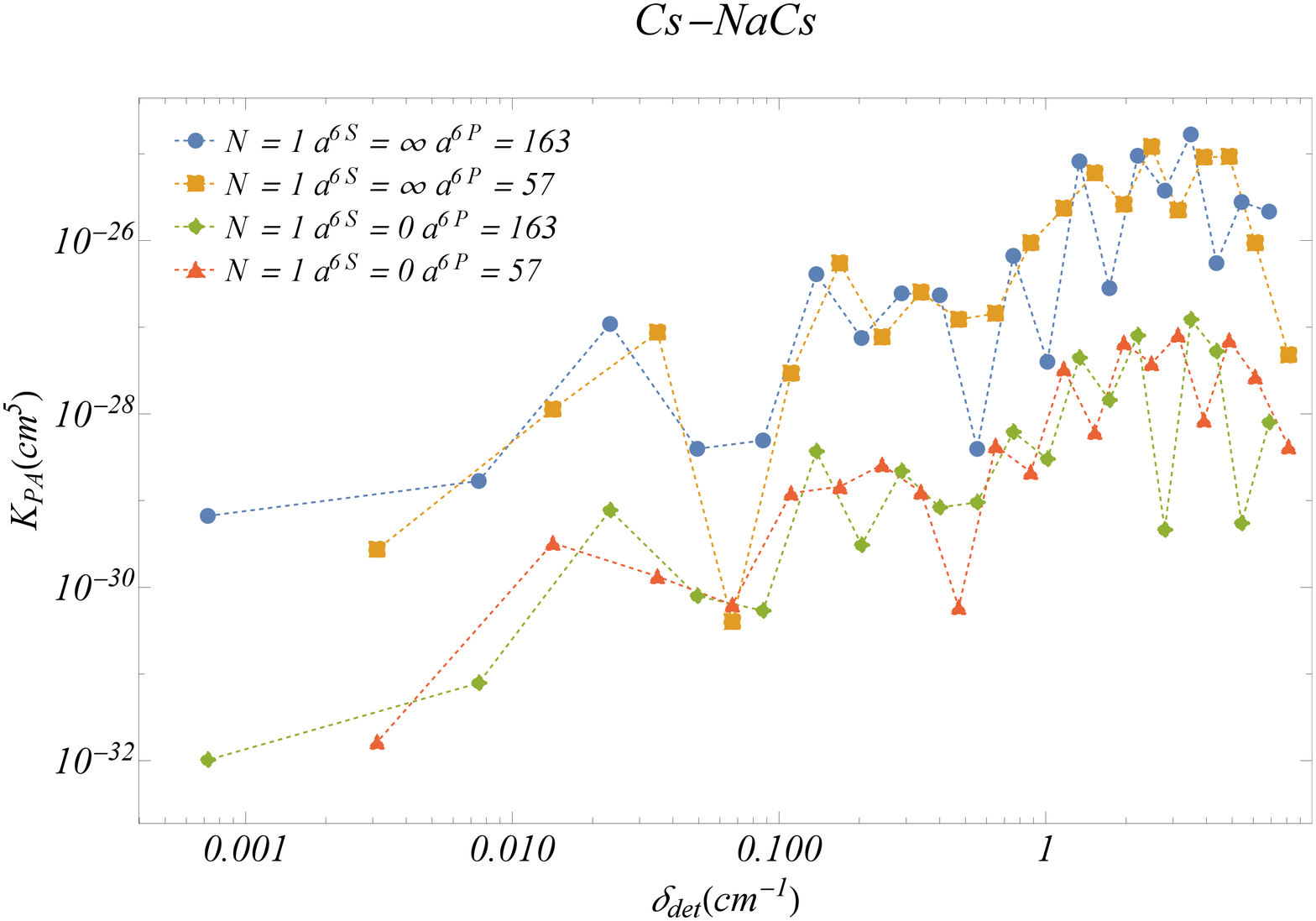}}
    \subfigure[]{\includegraphics[width=1\columnwidth]{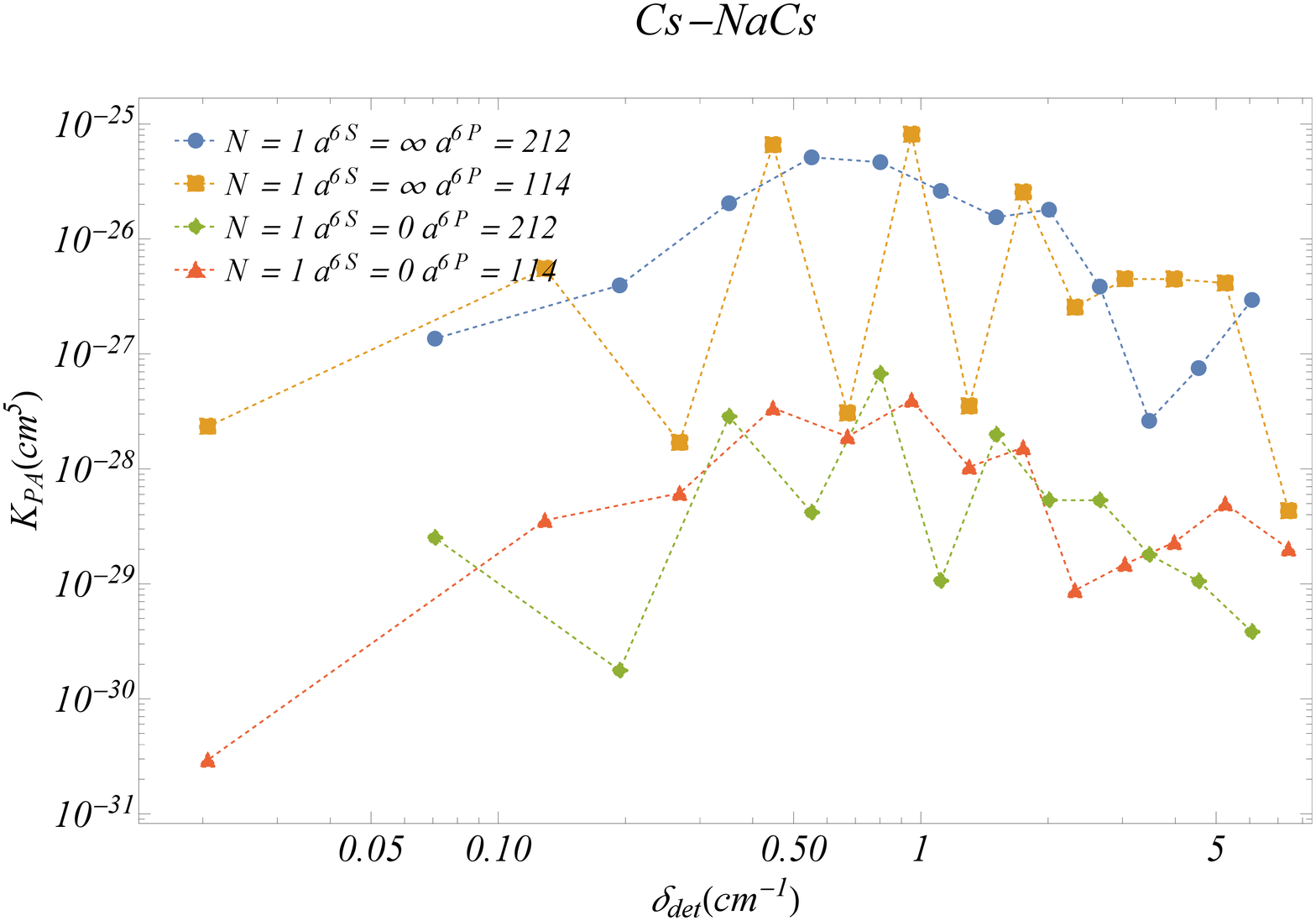}}
    \caption{The normalized PA rate $K_{PA}$ is plotted for different vibrational states. The initial state is $\ket{S;N = 0;E}$ and final states with $\ket{P_{3/2};N = 1;\epsilon_{\nu}<0;|\omega| = 1/2}$ for Cs-NaCs at average collision energy $T = 200 nK$. Three curves have $\ket{N = 1;|\omega| = 1/2}$ that are adiabatically traced from the lowest energy (a) to the highest energy(c). The different colors on each figure  are for the different boundary conditions considered in this study as shown in each inset.}
    \label{fig:PA_cs N = 1 w = 0.5}
\end{figure}

 \begin{figure}[t!]
    \centering
    \captionsetup{width=1\columnwidth}
    \subfigure[]{\includegraphics[width=1\columnwidth]{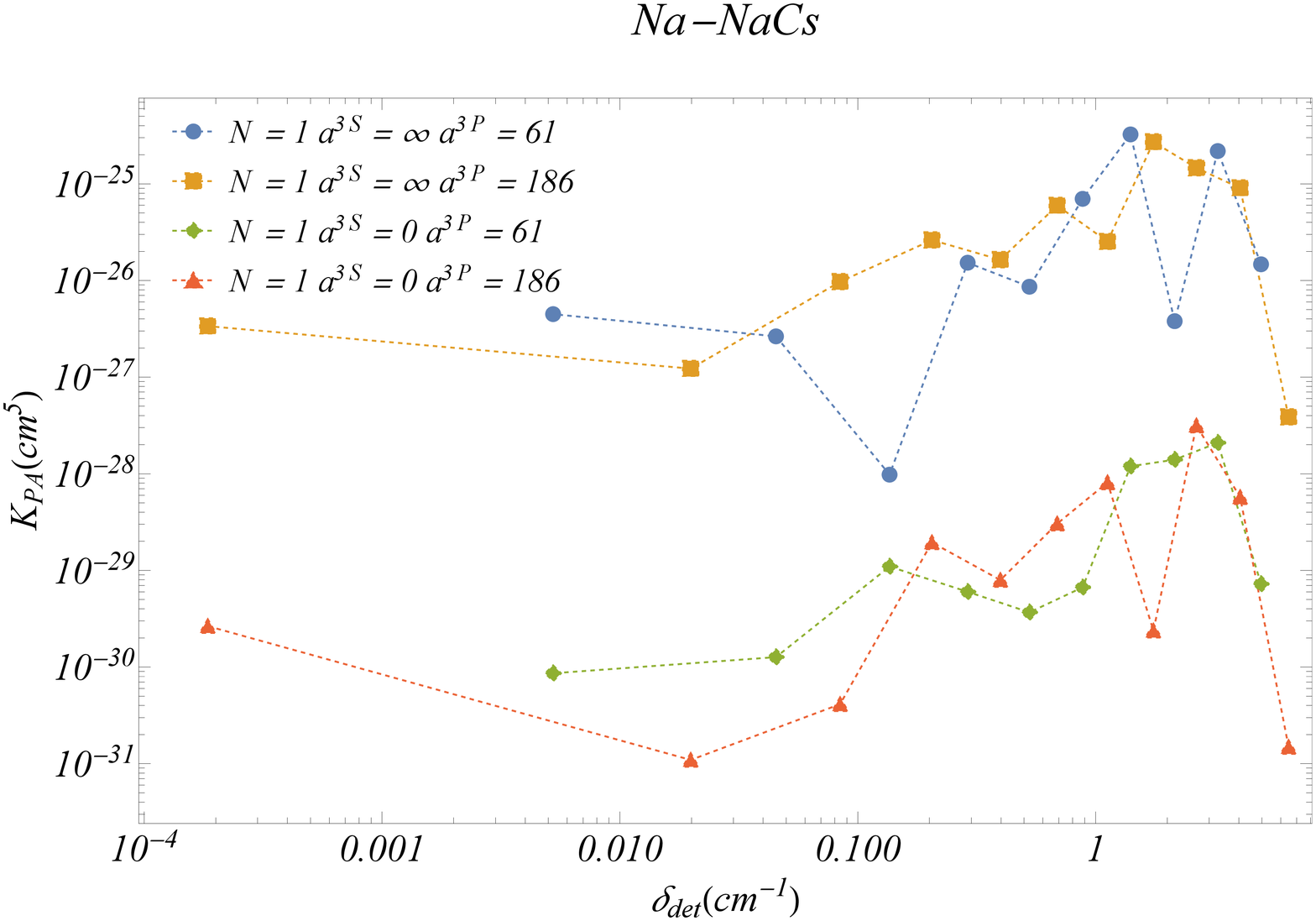}}
    \subfigure[]{\includegraphics[width=1\columnwidth]{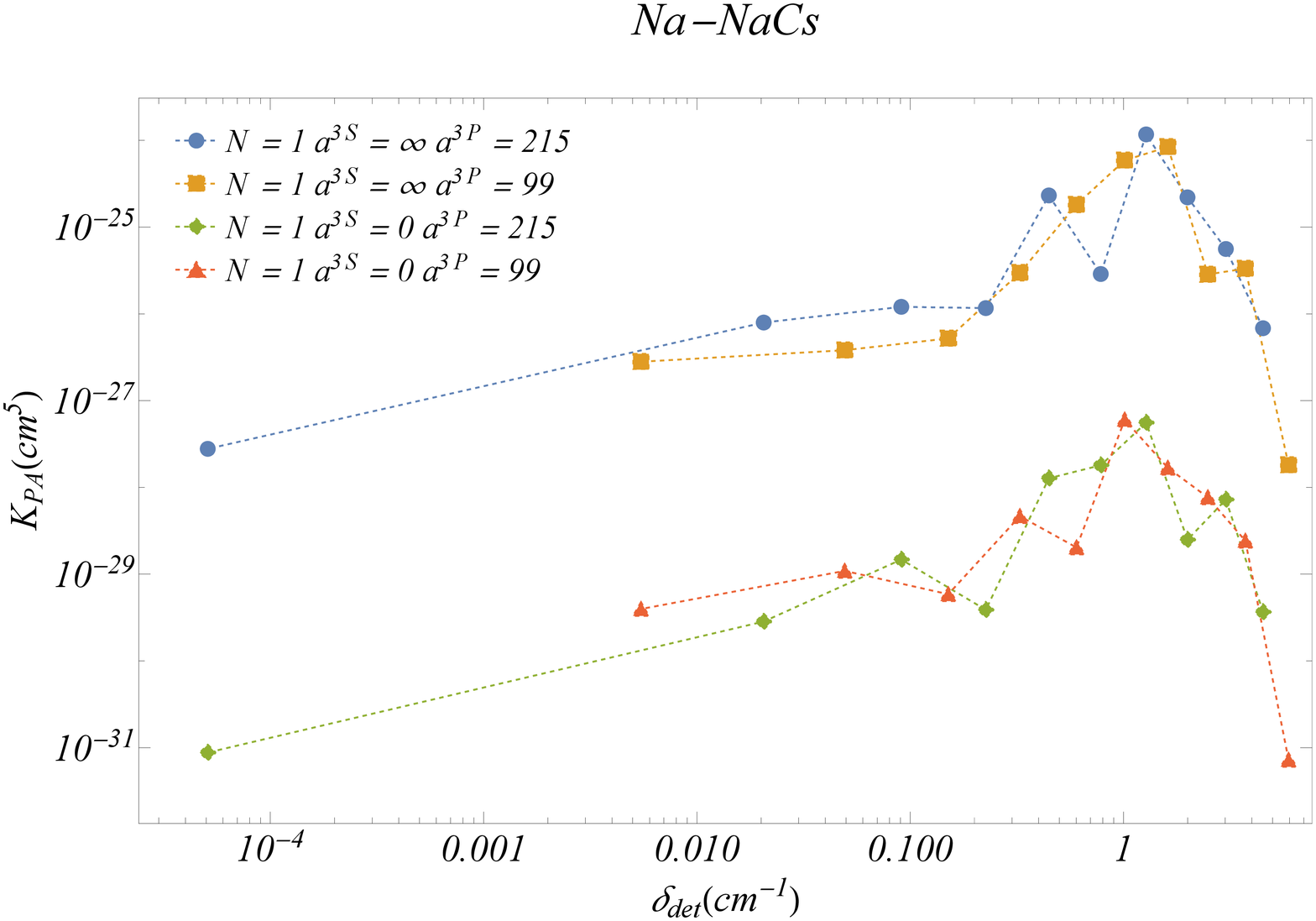}}
    \subfigure[]{\includegraphics[width=1\columnwidth]{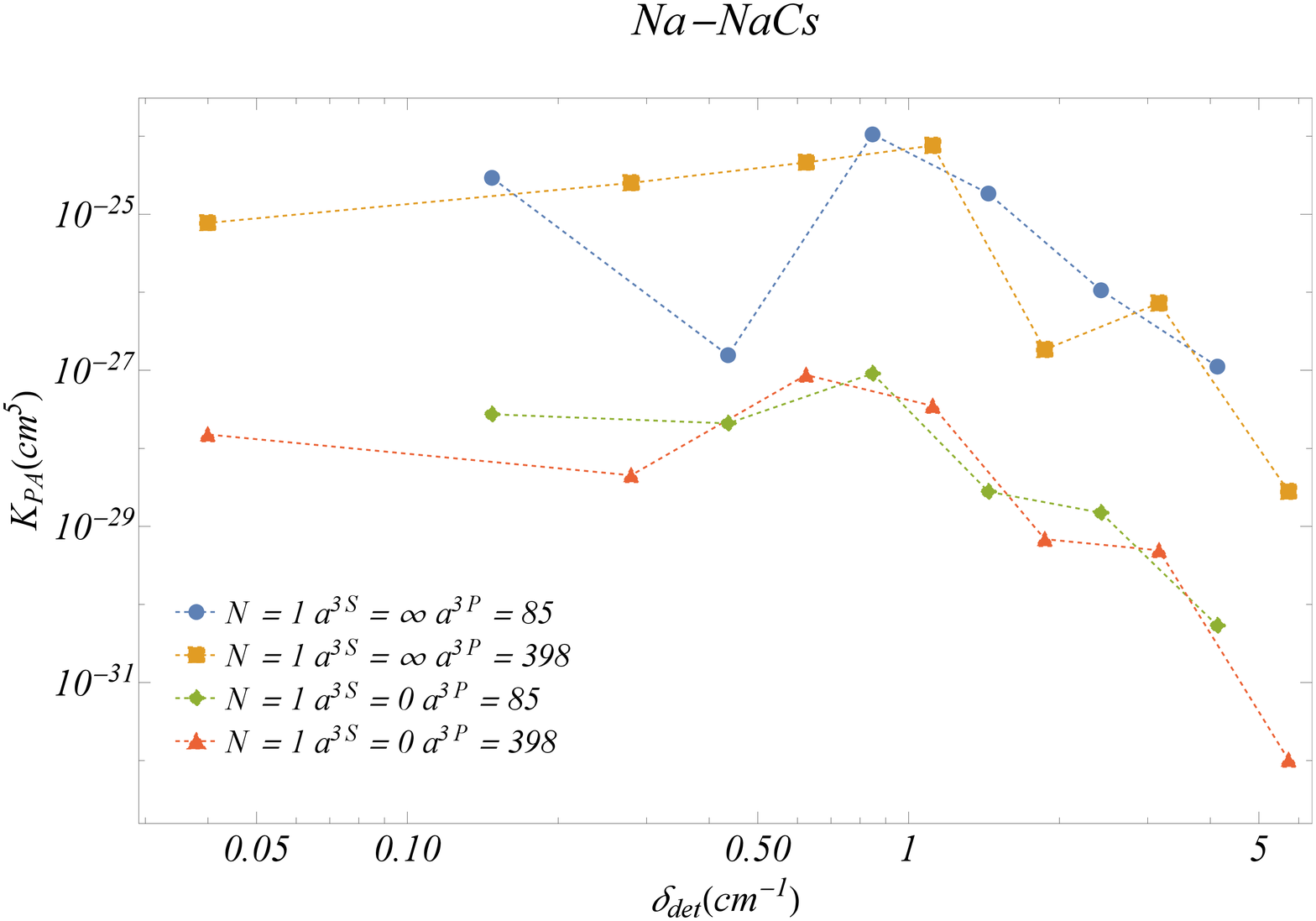}}
    \caption{Same as Fig.\ref{fig:PA_cs N = 1 w = 0.5} but for Na-NaCs}
    \label{fig:PA_na N = 1 w = 0.5}
\end{figure}

Finally, a brief derivation of Eq.\ref{atm_matrx_element} is presented. 
The spin-orbit coupled basis $(l,s)_{jm_j}$ are expanded according to
\begin{equation}\label{LS_coupled}
    \ket{(l,s)_{jm_j}} = \sum_{m_l,m_s} C_{lm_l,sm_s}^{jm_j}\ket{l,m_l}\ket{s,m_s}
\end{equation}
The multipole moment $Q_l^m$depends only on the space coordinates and does not affect the spin part. The wave function of the outer electron in an alkali atom is seperated into a radial part $u(r)/r$ and and angular part $\ket{(l,s)_{jm_j}}$. The $r$ dependence of $Q_l^m$ comes in $r^l$, so the radial contribution of the matrix element in Eq.\ref{atm_matrx_element} always has the form $\mathcal{A}_{n,l,j}^{n',l',j'} \equiv \int_0^{\infty} dr u_{n l j}(r)r^Lu_{n'l'j'}(r)$. Now Eq.\ref{LS_coupled} is used to evaluate
\begin{multline}\label{expans}
    \braket{n',l',j',mj'}{Q_L^M|n,l,j,mj} = \mathcal{A}_{n,l,j}^{n',l',j'}\sum_{m_l,m_s,m_l'}\\ 
    \times C_{l'm_l',s'm_s'}^{j'm_j'}C_{lm_l,sm_s}^{jm_j}\braket{l',m_l'}{\mathcal{Q}_L^M|l,m_l}
\end{multline}
 where $\mathcal{Q}_L^M = Q_L^M/(r^{L}) = \sqrt{\frac{4\pi}{2L+1}}Y_{L,M}$, a purely angular operator whose matrix element is given by\cite{varsh}
 \begin{multline}
     \braket{l',m_l'}{\mathcal{Q}_L^M|l,m_l} =  \sqrt{\frac{4\pi}{2L+1}}\braket{Y_{l',m_l'}}{Y_{L,M}|Y_{l,m_l}} \\
      = \sqrt{\frac{2l+1}{2l'+1}}C_{l0,L0}^{l'0}C_{lm_l,LM}^{l'm_l'}
 \end{multline}
We use the properties of the Clebsch-Gordan coefficients to evaluate the sum \cite{varsh}
 \begin{multline}\label{simpl}
     \sum_{m_l,m_s,m_l'} C_{l'm_l',sm_s}^{j'm_j'} C_{lm_l,sm_s}^{jm_j} C_{lm_l,LM}^{l'm_l'} \\ 
      = (-1)^{l'+s+j+L}\sqrt{(2l'+1)(2j+1)} C_{jm_j,LM}^{j'm_j'} \\
      \times \begin{Bmatrix}
l & s & j \\
j' & L & l'
\end{Bmatrix}\\
\end{multline}

Now,  combining Eq.\ref{simpl} with\ref{expans}, matrix element is evaluated and has the same expression used in Eq.\ref{atm_matrx_element}.
\begin{multline}
\braket{n',l',j',mj'}{Q_L^M|n,l,j,mj} =-e \mathcal{A}_{n,l,j}^{n',l',j'}\\
\times (-1)^{j+L+l'+1/2}  \sqrt{(2j+1)(2l+1)} \\
 \times \sqrt{\frac{4\pi}{2L+1}}C_{j m_j,L M}^{j' m_j'}C_{l 0,L 0}^{l' 0}\begin{Bmatrix}
l & s & j \\
j' & L & l'
\end{Bmatrix}
\end{multline}
\newpage
%%%%%%%%%%%%%%%%%%%

%%%%%%%%%%%%%%%%%%
 \begin{figure}[H]
\centering
\captionsetup{width=1\columnwidth}
\subfigure[]{\includegraphics[width=1\columnwidth]{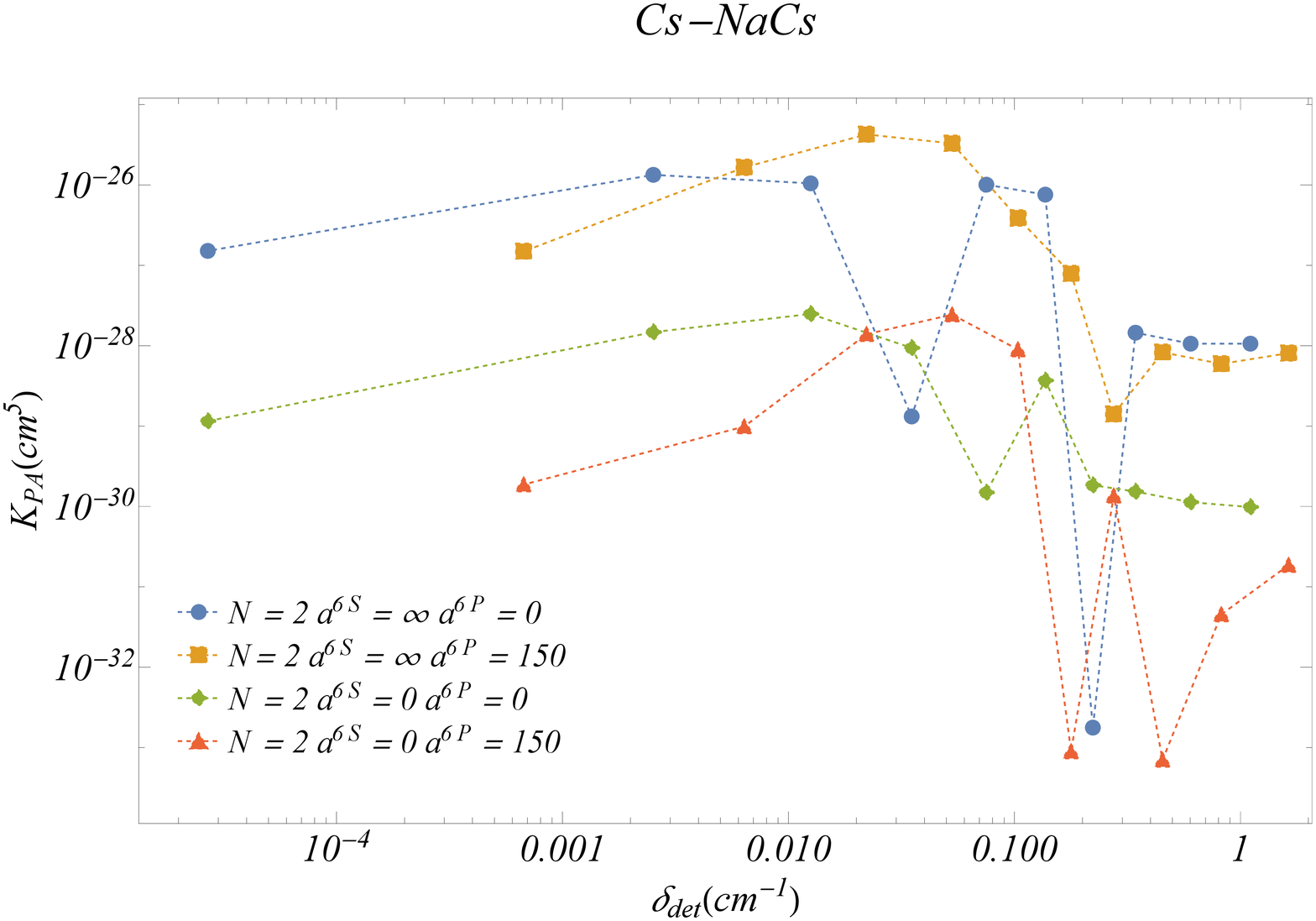}}
\subfigure[]{\includegraphics[width=1\columnwidth]{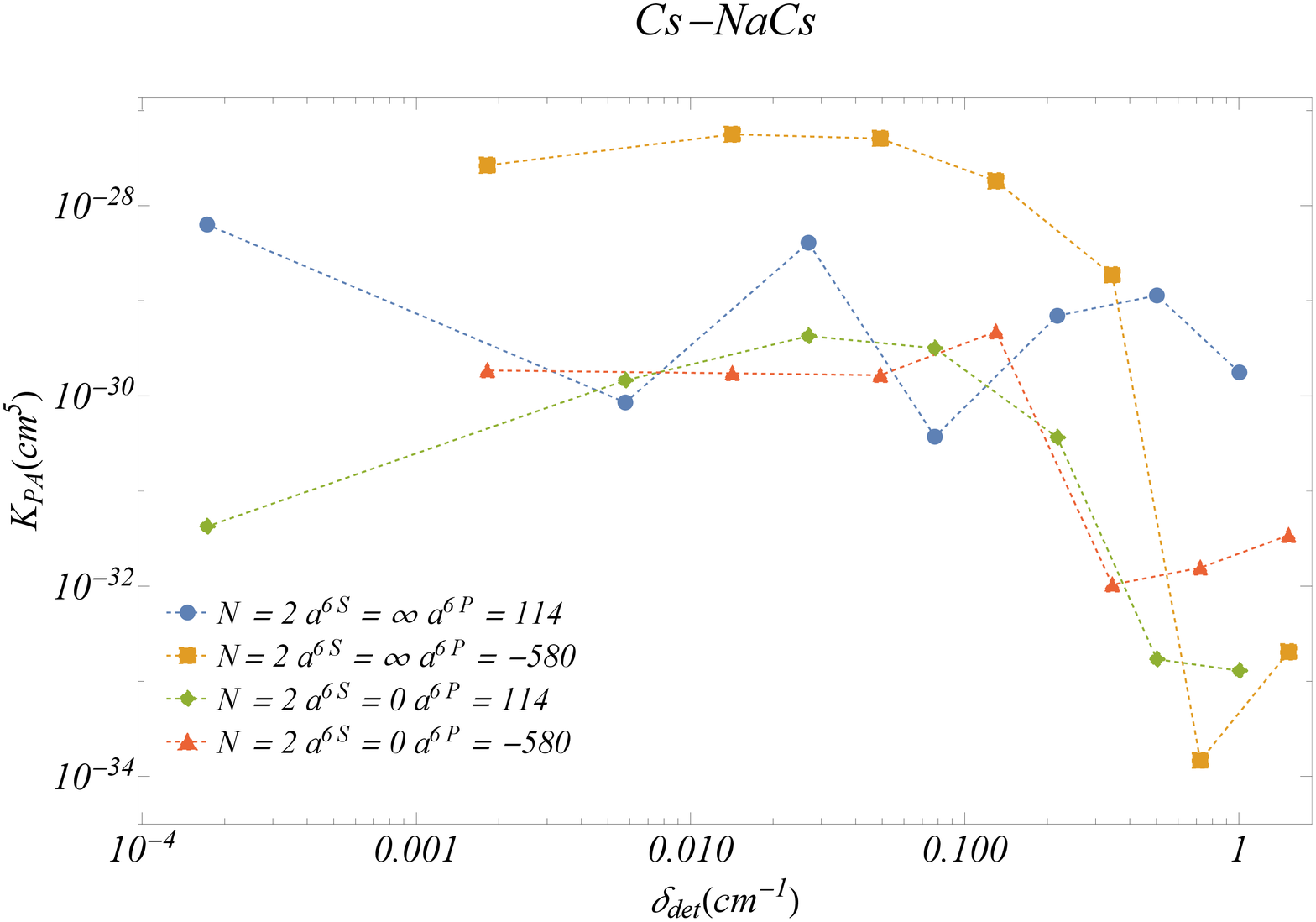}}
\subfigure[]{\includegraphics[width=1\columnwidth]{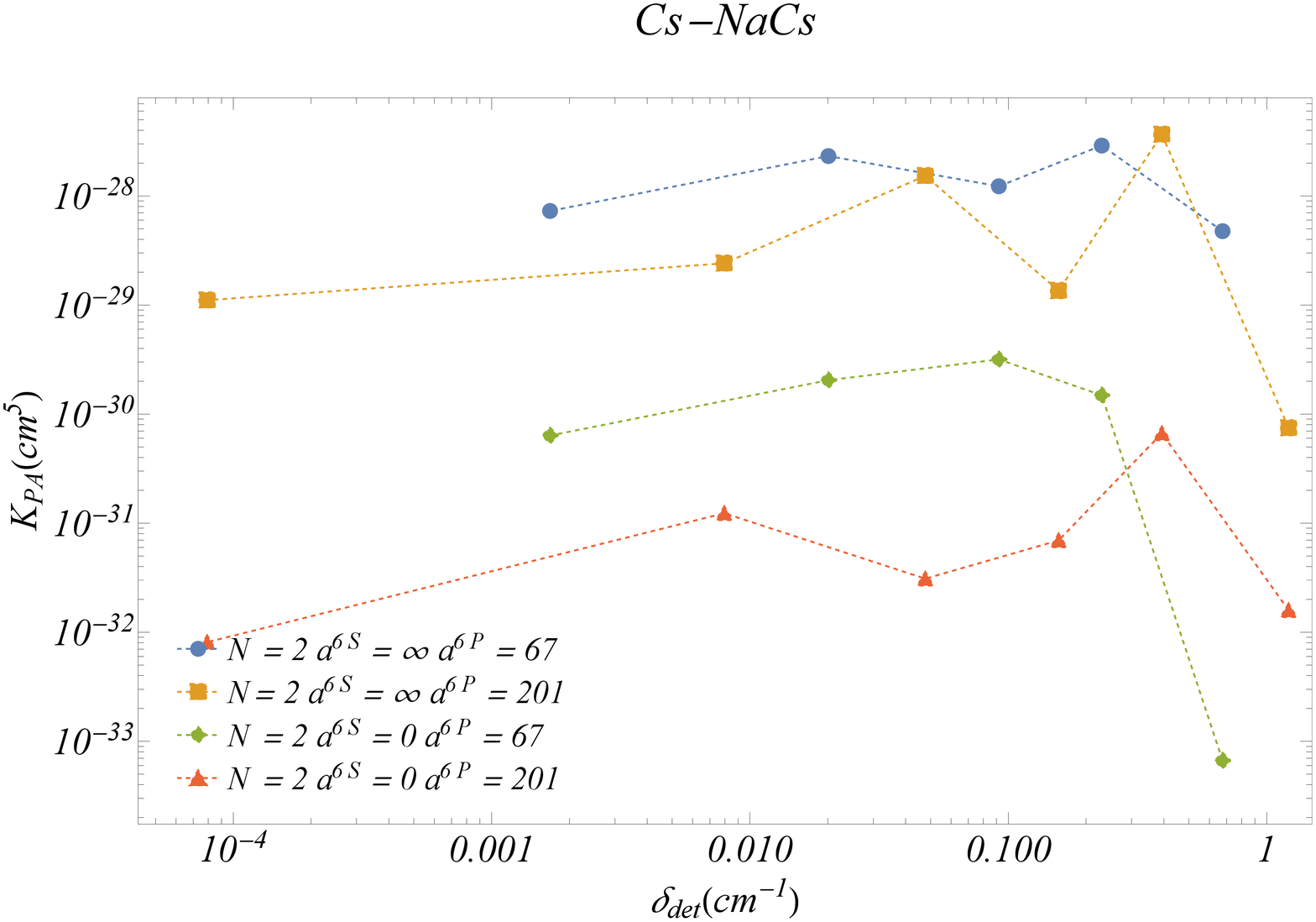}}
\caption{The normalized PA rate $K_{PA}$ for Cs-NaCs is plotted for different vibrational states, each with the symmetry $\ket{P_{3/2};N = 2;\epsilon_{\nu}<0;|\omega| = 1/2}$. The initial state is $\ket{S;N = 0;E}$ with average collision energy $T = 200 nK$. Four curves have $\ket{N = 2;|\omega| = 3/2}$ that are adiabatically traced from the lowest energy (a) to the highest energy(c). The different colors on each figure  are for the different boundary conditions considered in this study as shown in each inset.}
\label{fig:PA_cs N = 2 w = 1.5}
\end{figure}
%%%%%%%%%%%%%%%%%%%
\begin{figure}[H]
\centering
\captionsetup{width=1\columnwidth}
\subfigure[]{\includegraphics[width=1\columnwidth]{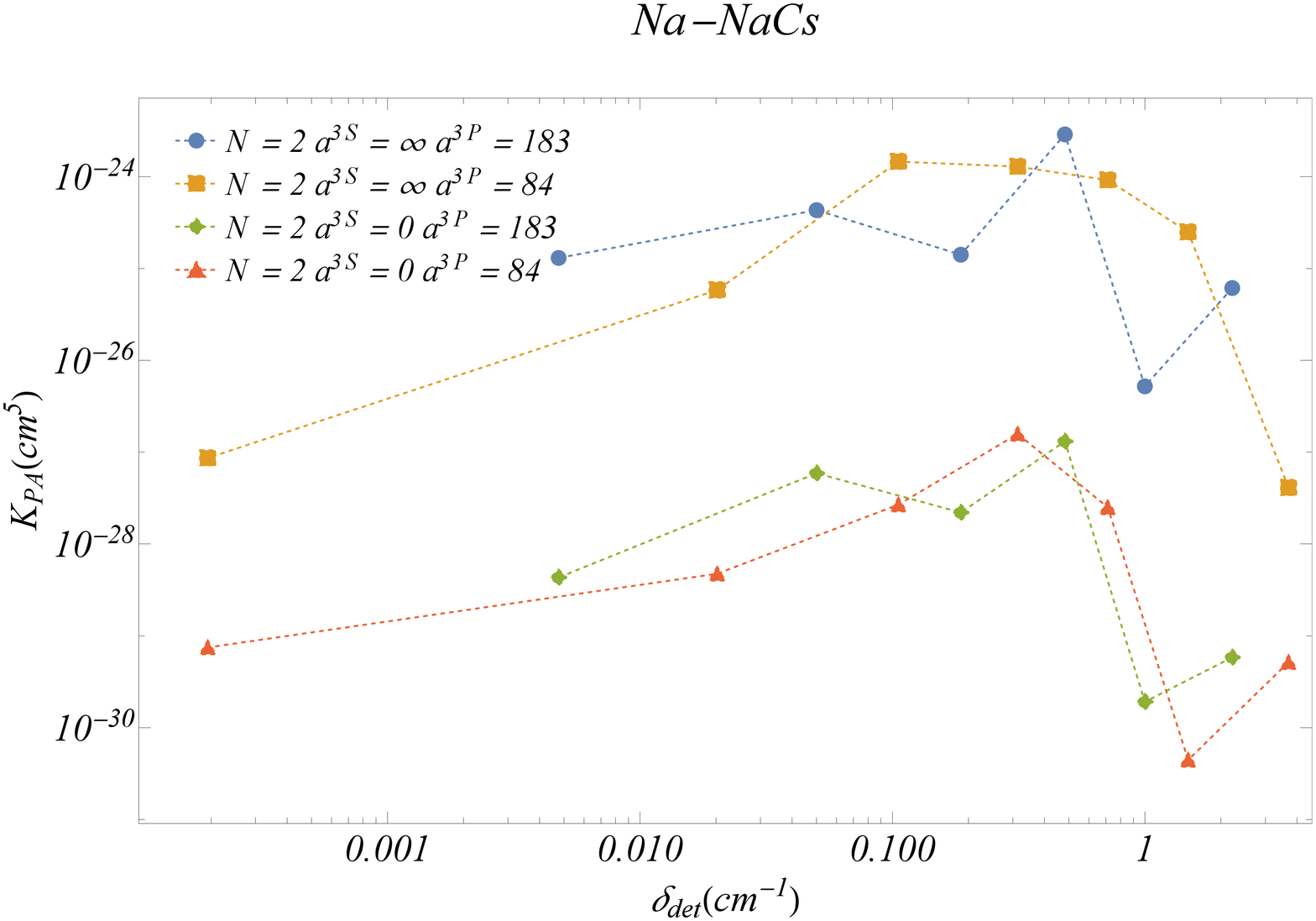}}
\subfigure[]{\includegraphics[width=1\columnwidth]{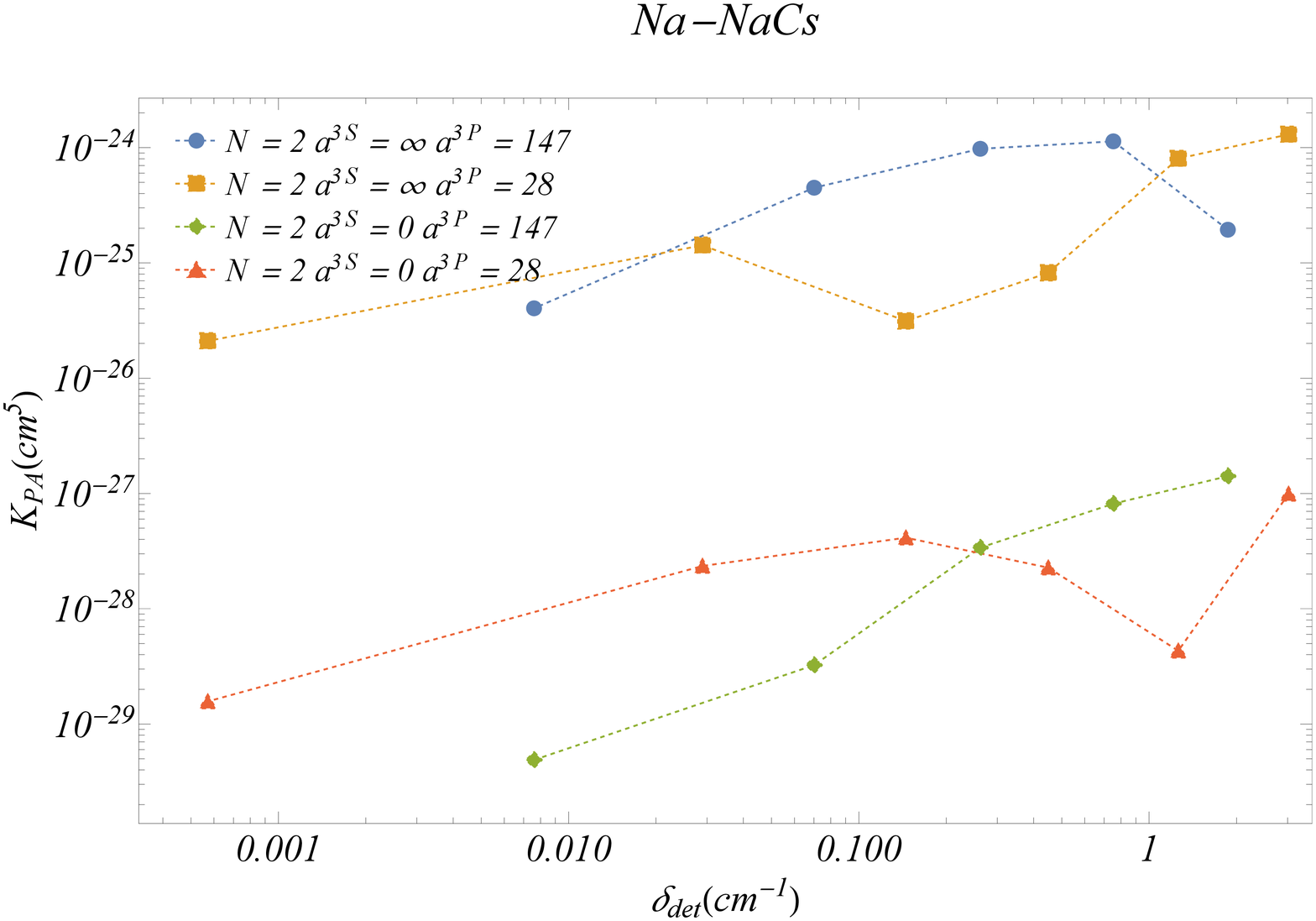}}
\subfigure[]{\includegraphics[width=1\columnwidth]{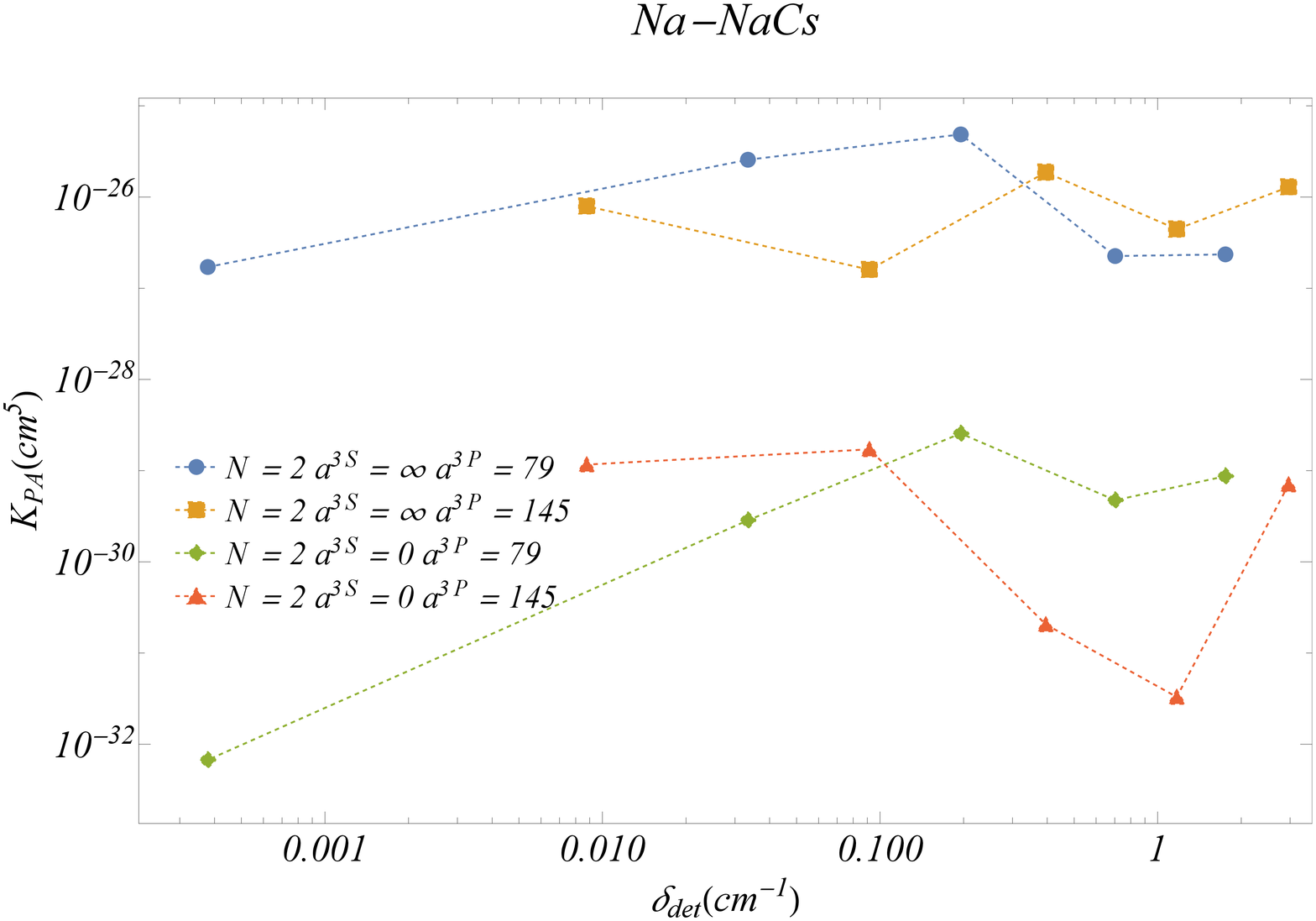}}
\caption{Same as Fig.\ref{fig:PA_cs N = 2 w = 1.5} but for Na-NaCs}
\label{fig:PA_na N = 2 w = 1.5}
\end{figure}
\newpage

%%%%%%%%%%%%%%%%%%%%%%%%
\begin{figure}[htp]
    \centering
    \captionsetup{width=1\columnwidth}
    \subfigure[]{\includegraphics[width=1\columnwidth]{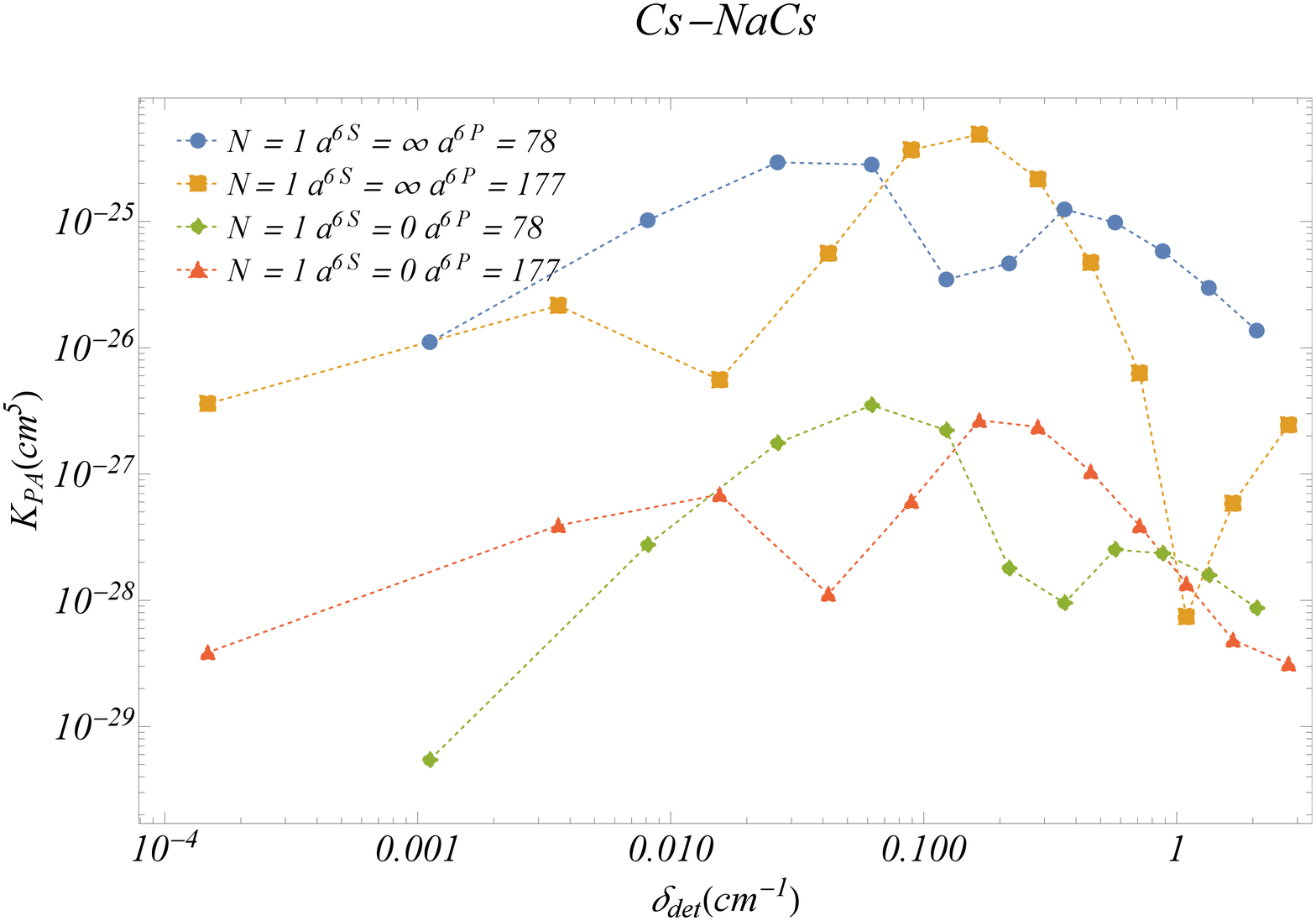}}
    \subfigure[]{\includegraphics[width=1\columnwidth]{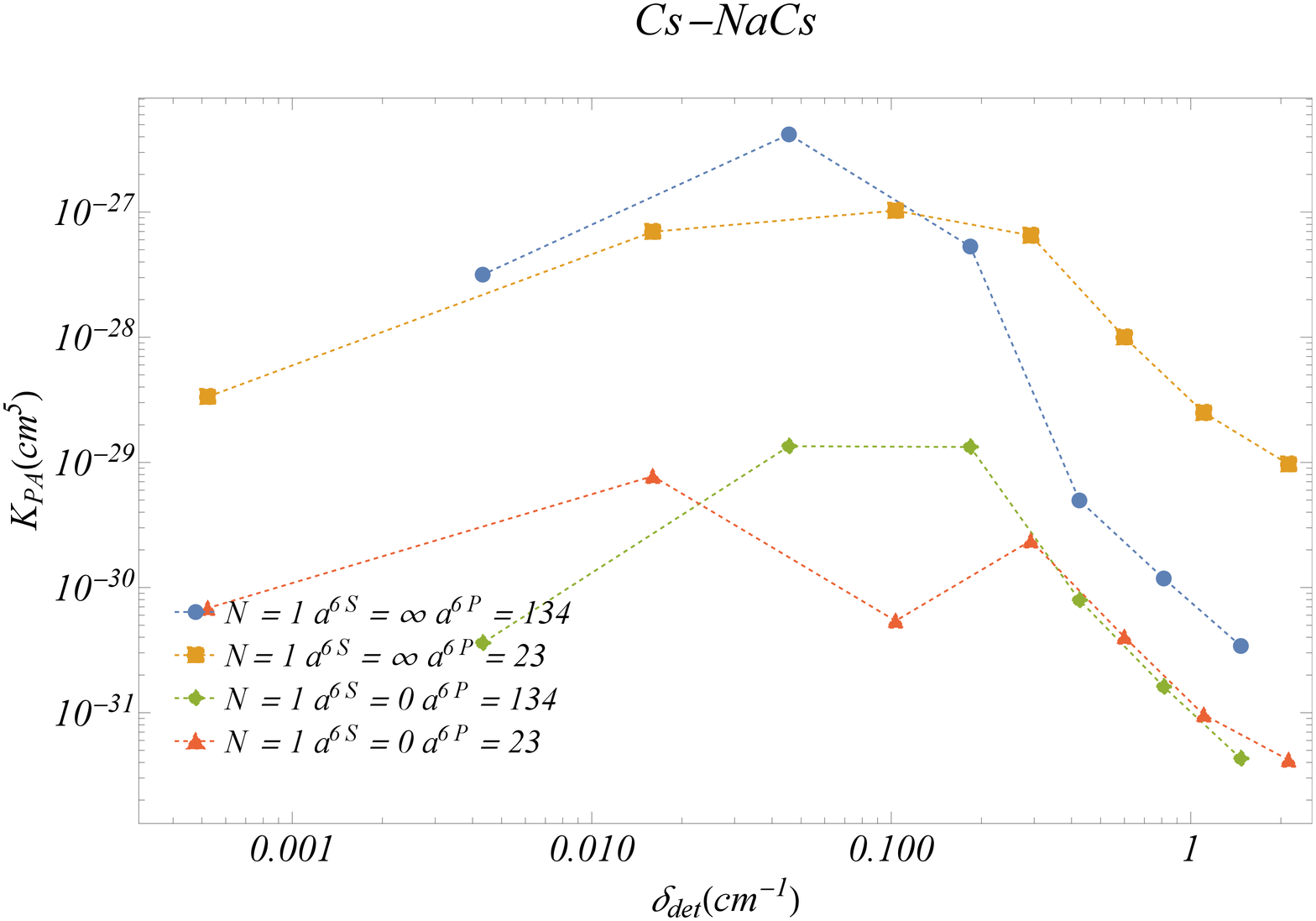}}
    \caption{(a), and (b) The normalized PA rate $K_{PA}$ is plotted between initial state $\ket{S;N = 0;E}$ and final states with $\ket{P_{3/2};N = 1;\epsilon_{\nu}<0;|\omega| = 3/2}$ for Cs-NaCs at average collision energy $T = 200 nK$. Two curves have $\ket{N = 1;|\omega| = 3/2}$ that are adiabatically traced from the lowest energy (a) to the highest energy(b).}
    \label{fig:PA_cs N = 1 w = 1.5}
\end{figure}
%%%%%%%%%%%%%%%%%%%%%%%%%%%%%%%%

\begin{figure}[t!]
    \centering
    \captionsetup{width=1\columnwidth}
    \subfigure[]{\includegraphics[width=1\columnwidth]{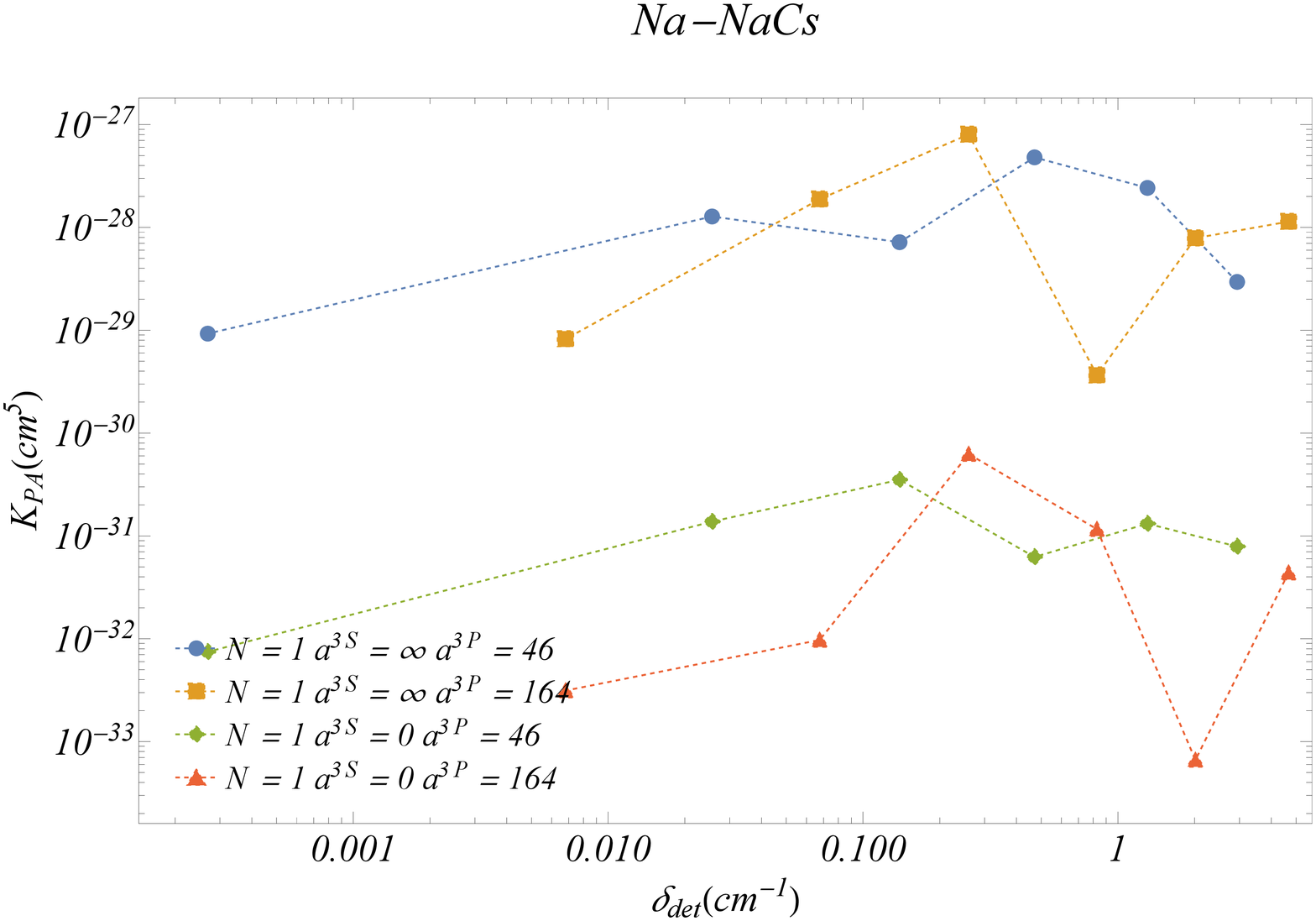}}
    \subfigure[]{\includegraphics[width=1\columnwidth]{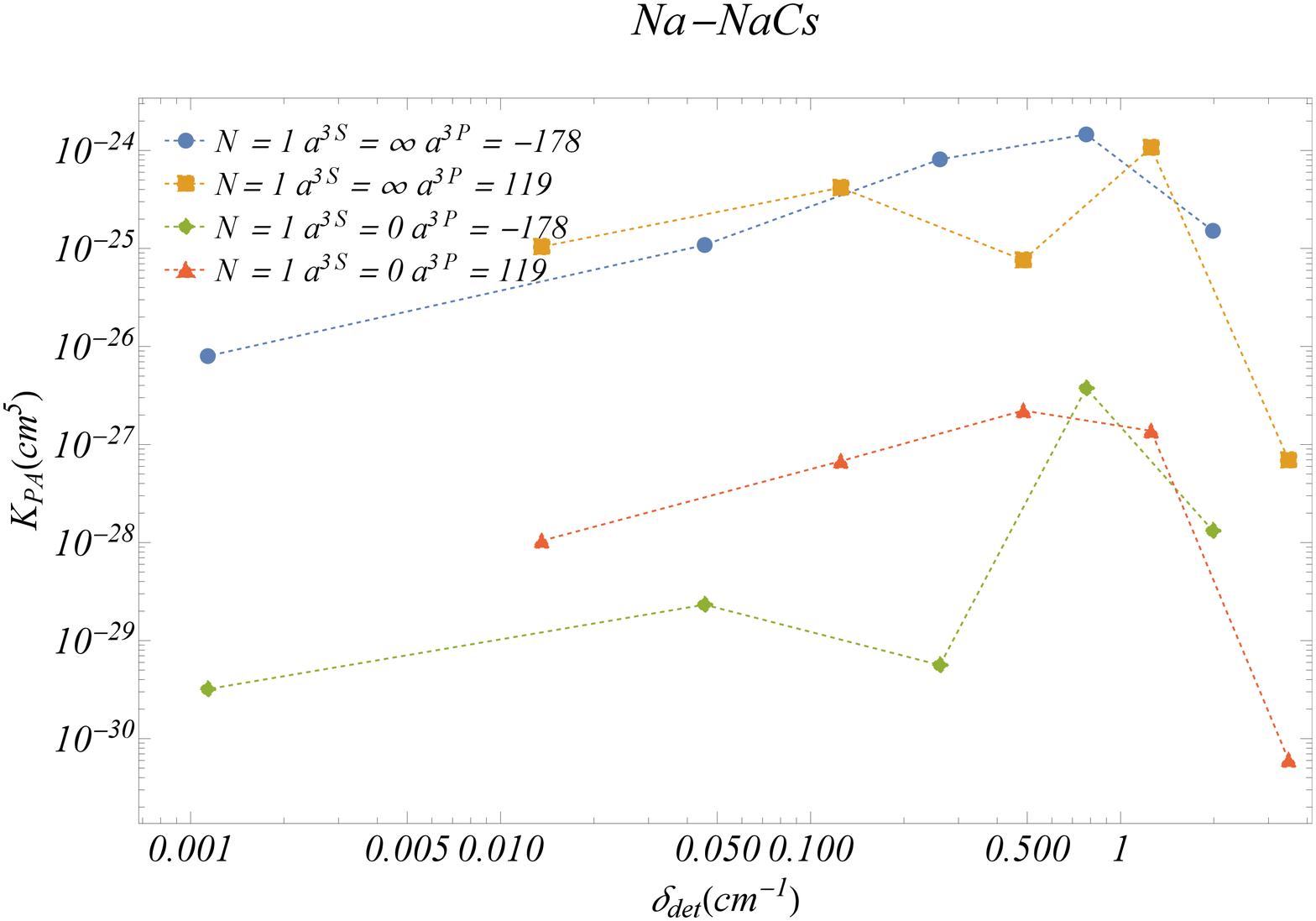}}
    \caption{same as Fig\ref{fig:PA_cs N = 1 w = 1.5} but for Na-NaCs}
    \label{fig:PA_na N = 1 w = 1.5}
\end{figure}

\section{Photoassociation to higher rotational states}
\label{sec:appendixB}
In Sec.\ref{subsec:level2}, PA rate is calculated and shown for different final states, all with the lowest rotational quantum number of the dimer $N = 0$. In this section, we show PA rate for different final states with non-zero rotational quantum number $N$. Since the matrix element in Eq.\ref{PA_rate} is for the atomic dipole, one expects the dipole transitions to be maximum between initial and final state both with the same $N$. Thus, the values of the PA rate for final states with non-zero $N$ are expected to be smaller.

\begin{figure*}[htp!]
\centering
%\captionsetup{width=2.0\columnwidth}
\subfigure[]{\includegraphics[width=1\columnwidth]{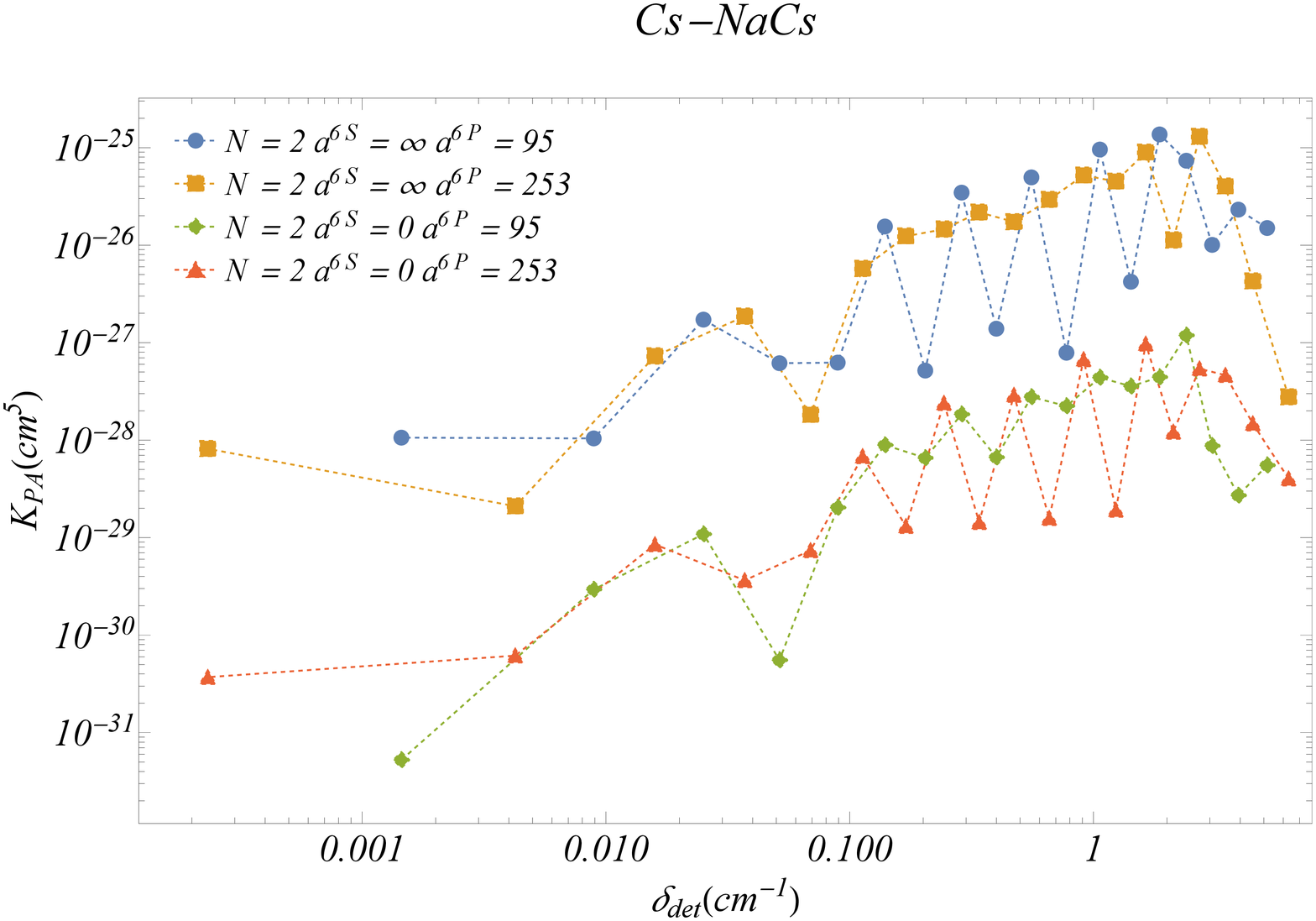}}\subfigure[]{\includegraphics[width=1\columnwidth]{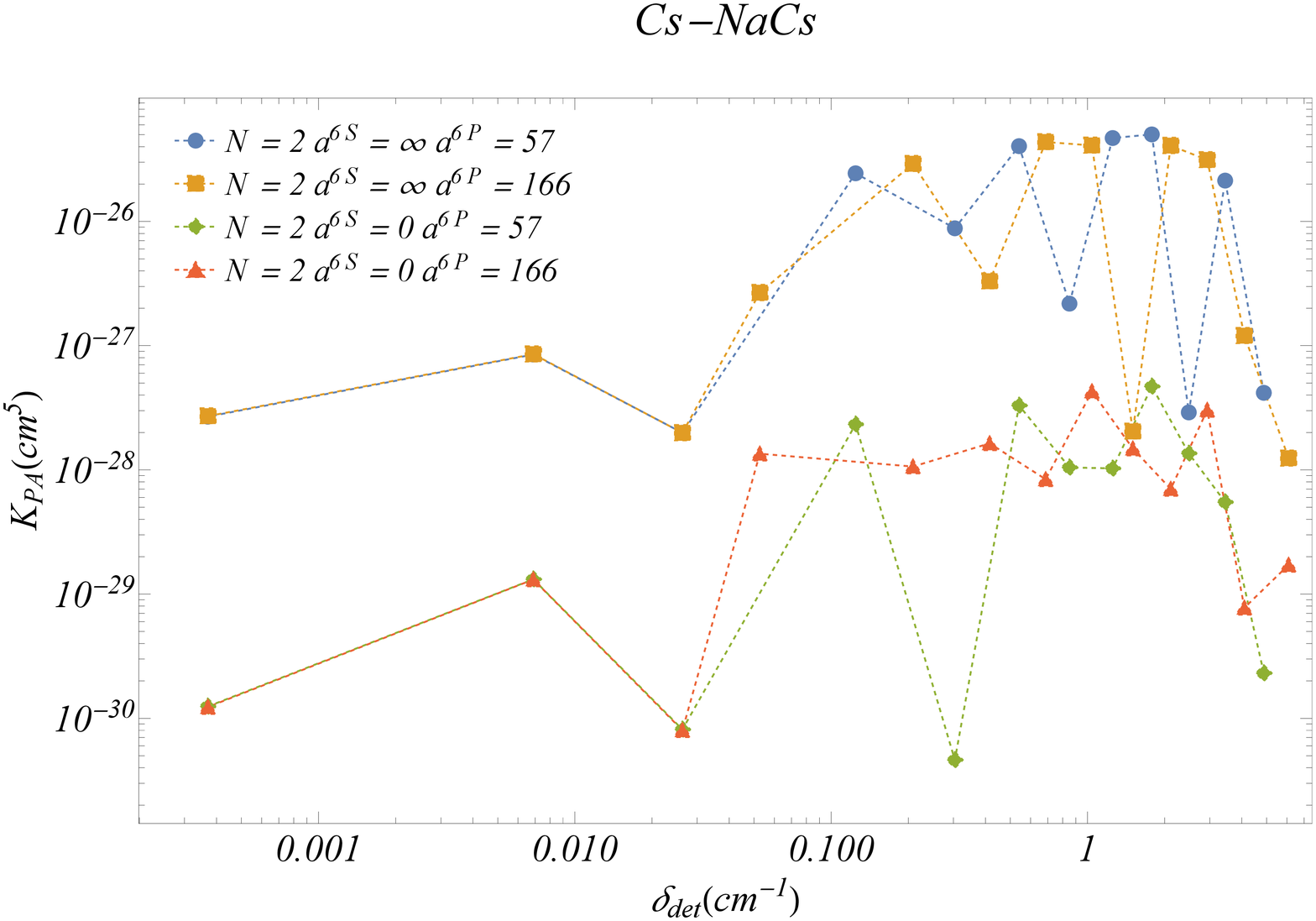}}
\subfigure[]{\includegraphics[width=1\columnwidth]{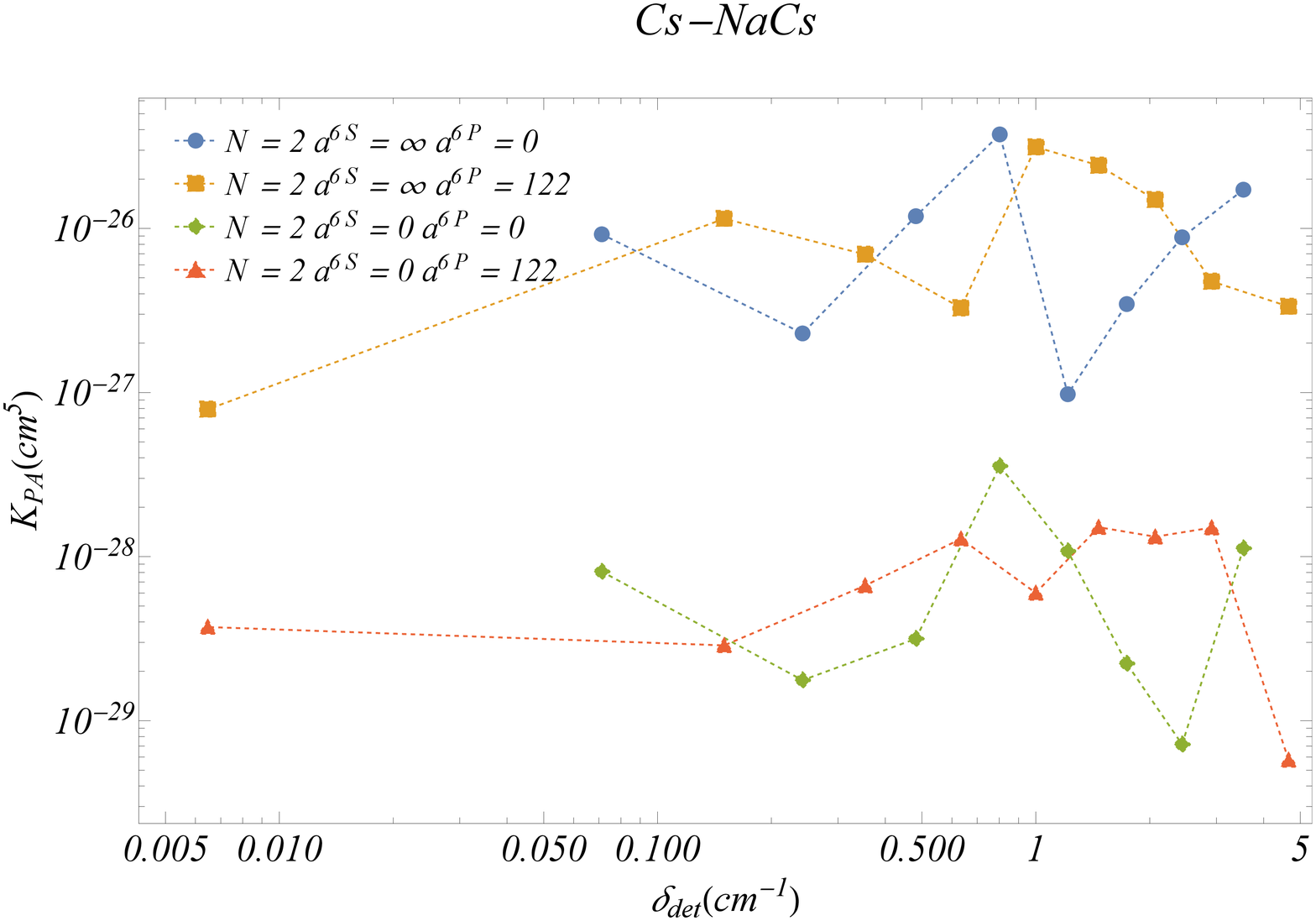}}\subfigure[]{\includegraphics[width=1\columnwidth]{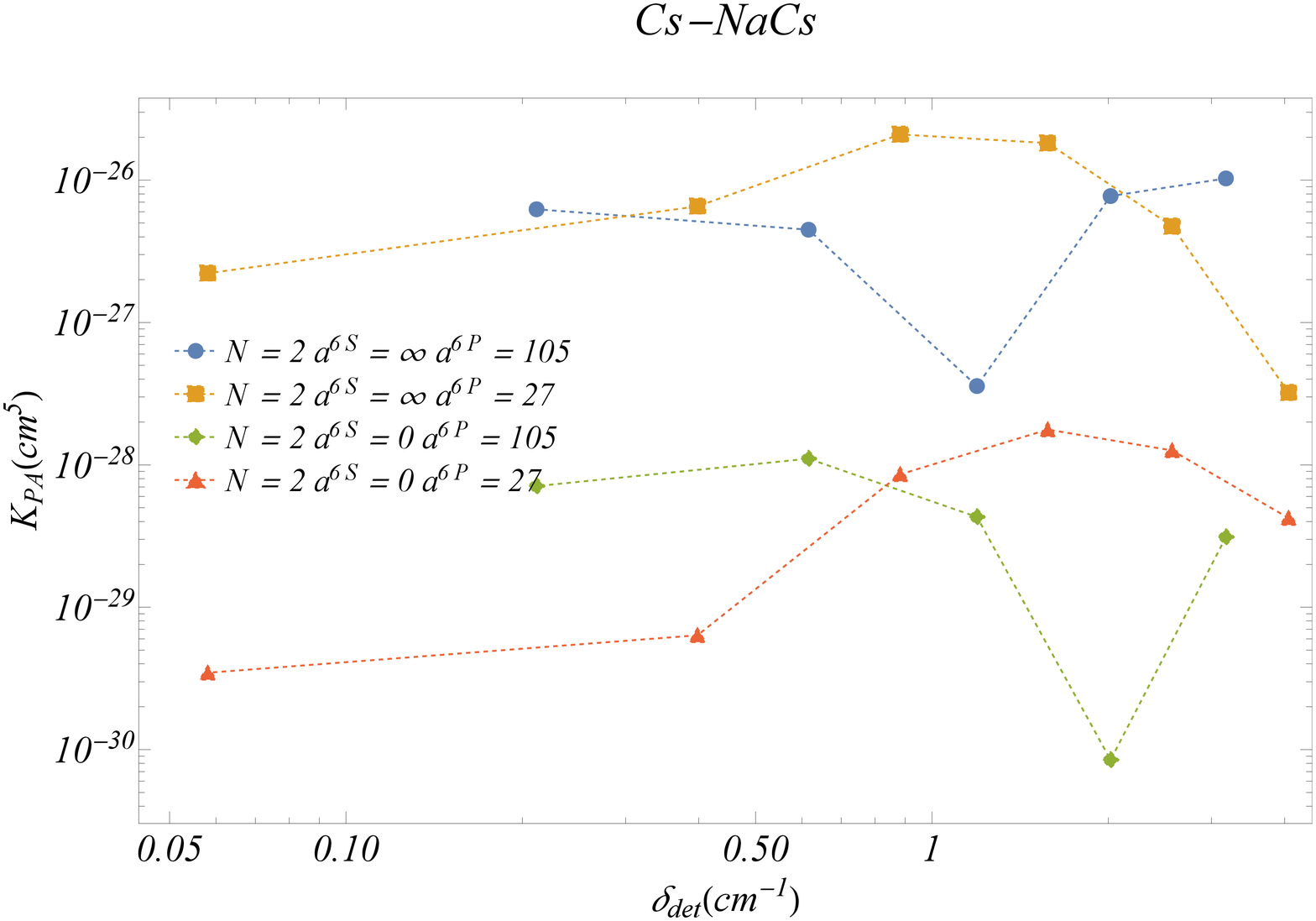}}
\caption{The normalized PA rate $K_{PA}$ for Cs-NaCs system is plotted for different vibrational states $\ket{P_{3/2};N = 2;\epsilon_{\nu}<0;|\omega| = 1/2}$. The Initial state of PA is $\ket{S;N = 0;E}$ at average collision energy $T = 200 nK$. Four curves have the symmetry $\ket{N = 2;|\omega| = 1/2}$ and are adiabatically traced from the lowest energy (a) to the highest energy(d). The different colors on each figure  are for the different boundary conditions considered in this study as shown in each inset.}
\label{fig:PA_cs N = 2 w = 0.5}
\end{figure*}

\begin{widetext}
\begin{figure}[H]
\centering
\captionsetup{width=1\columnwidth}
\subfigure[]{\includegraphics[width=1\columnwidth]{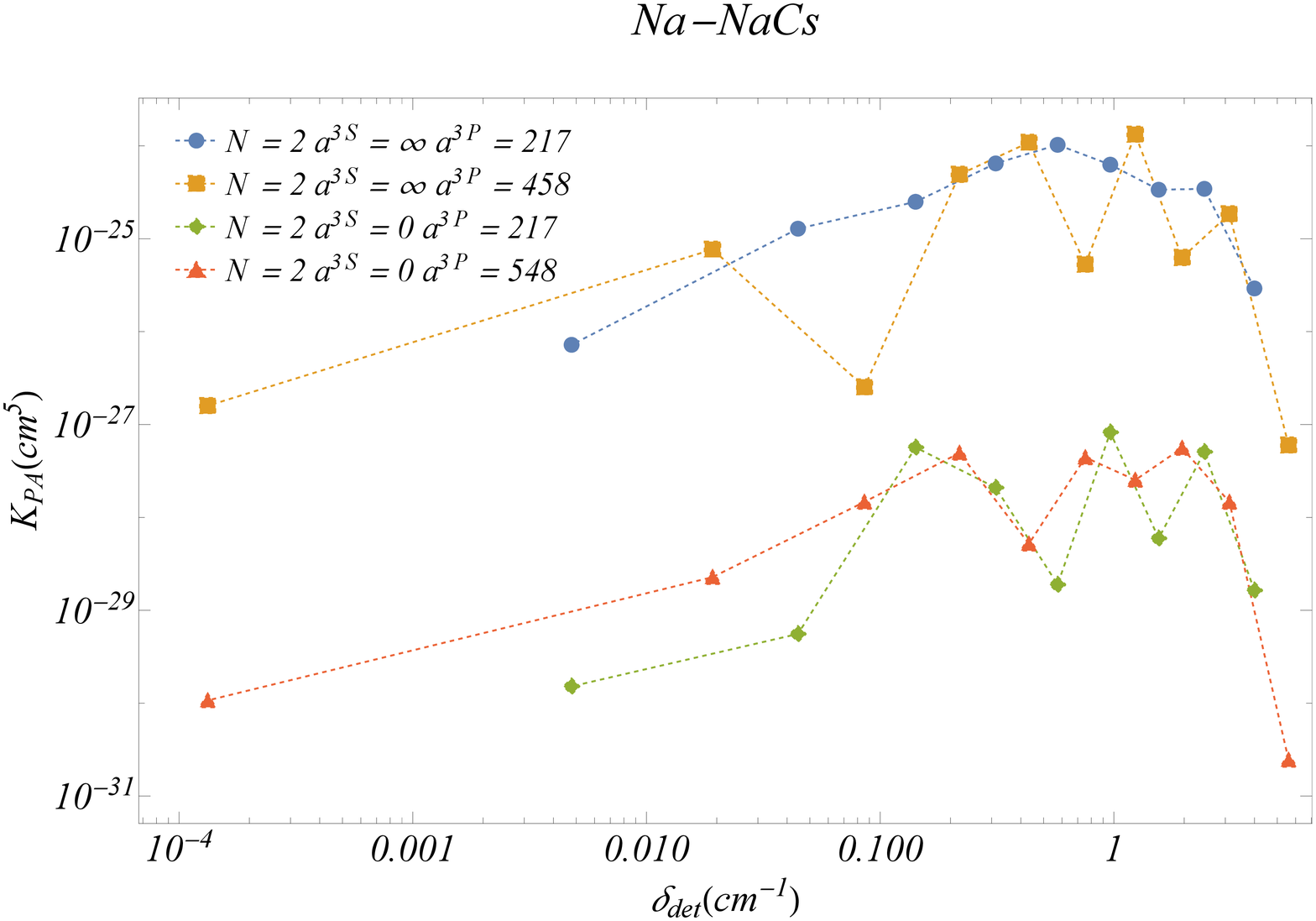}}\subfigure[]{\includegraphics[width=1\columnwidth]{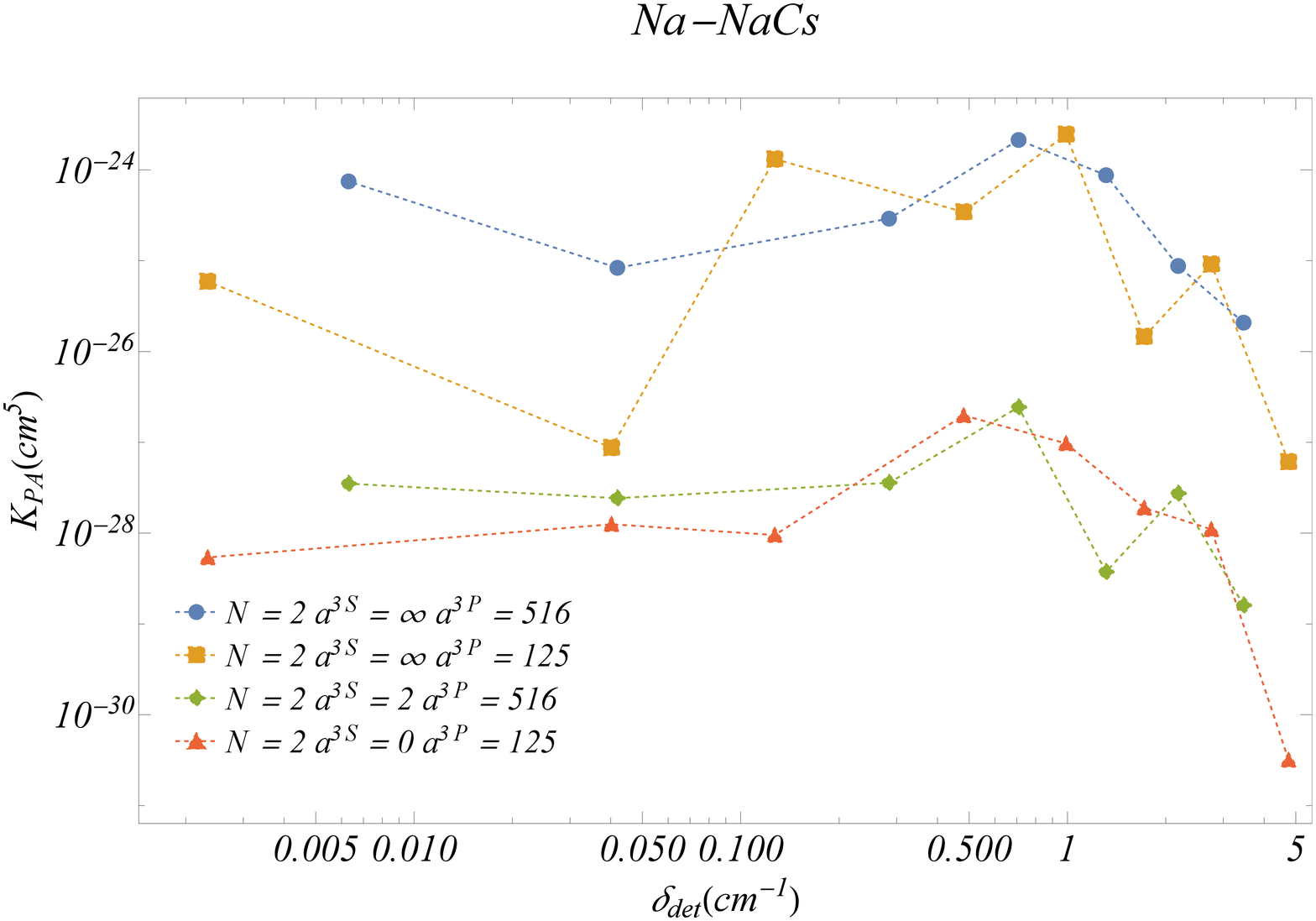}}
\subfigure[]{\includegraphics[width=1\columnwidth]{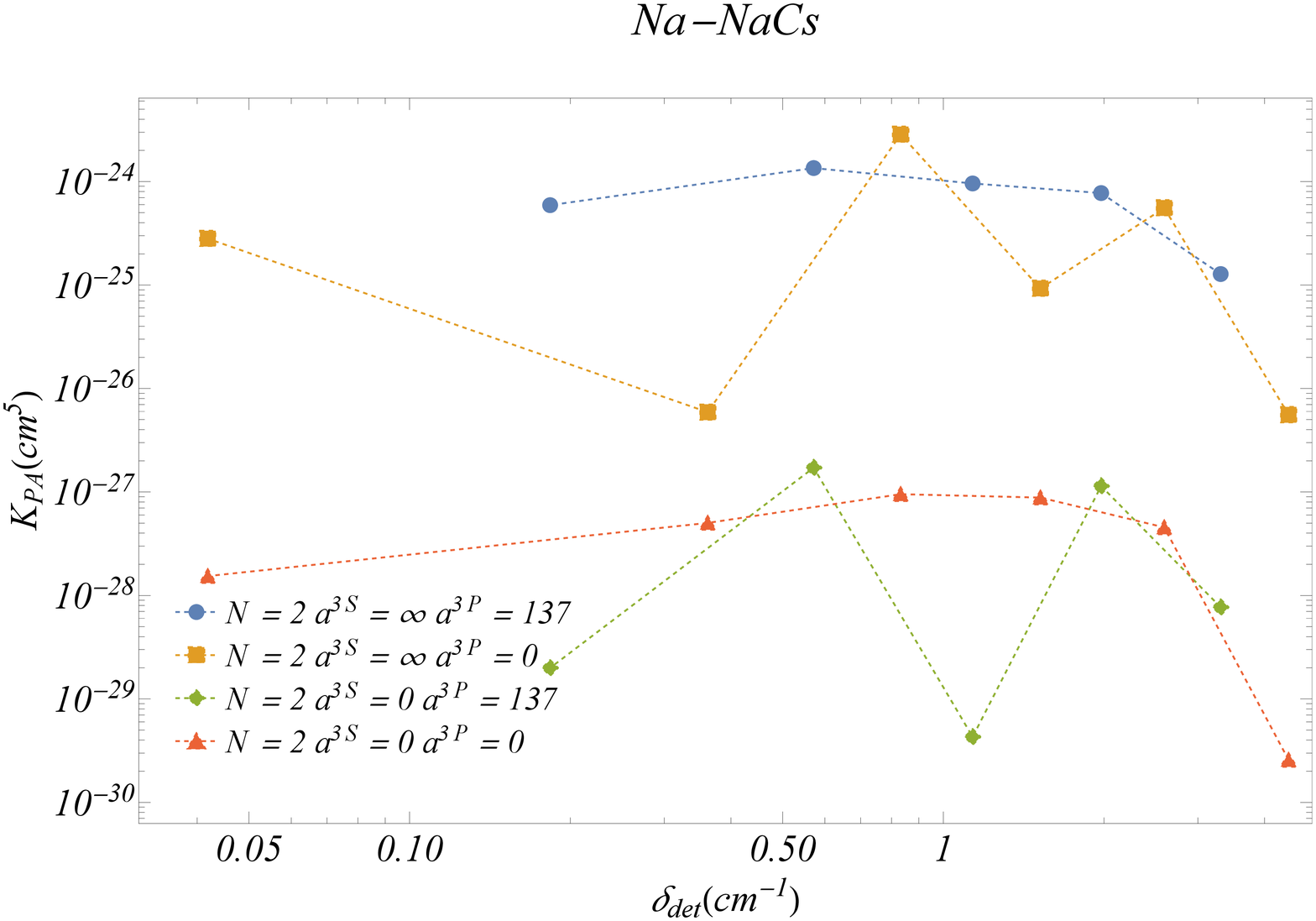}}\subfigure[]{\includegraphics[width=1\columnwidth]{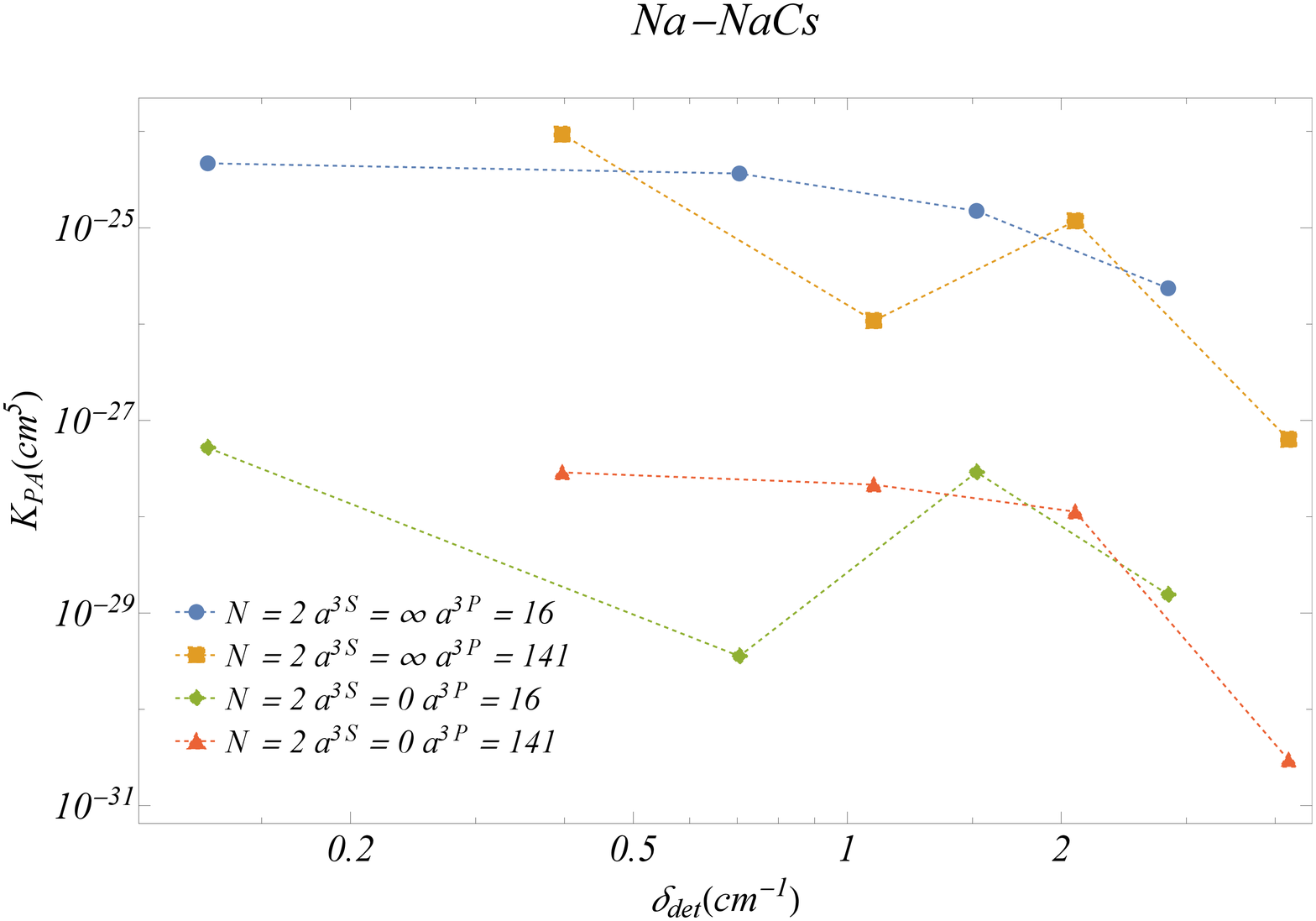}}
\caption{Same as Fig.\ref{fig:PA_cs N = 2 w = 0.5} but for Na-NaCs}
\label{fig:PA_na N = 2 w = 0.5}
\end{figure}
\end{widetext}

Throughout this article, we focused on dipole transition from the initial state which, at long distance, has quantum numbers $\ket{S;N = 0;|\omega| = 0.5}$. Such state couples to the final states $\ket{P_{3/2};N;|\omega| = (0.5,1.5)}$ for any value of $N$. The cases where $N = 0$ are shown in \ref{subsec:level2}. 
Fig.\ref{fig:PA_Na},\ref{fig:PA_CS}, \ref{fig:PA_wf = 1.5} and \ref{fig:PA_Cs3}. As for higher rotational levels ($N = 1,2$ and $|\omega| = 1/2$,$3/2$), figures \ref{fig:PA_cs N = 1 w = 0.5} $-$ \ref{fig:PA_na N = 2 w = 0.5} show the values of the PA rate for both systems of interest. The results in the present section and in Sec.\ref{subsec:level2} show that PA is suppressed for higher rotational levels $N$ with no long-range electric dipole coupling $\bra{\psi_i(E)}\vec{d}\ket{\psi_f(\nu)}$ to the ground state.

\newpage
%\begin{widetext}
%\end{widetext}
%%%%%%%%%%%%%%%%%%%%%%%%%%
%%%%%%%%%%%%%%%%%%%%%%%

\bibliographystyle{unsrt} % Use for unsorted references  
%\bibliographystyle{plainnat} % use this to have URLs listed in References
%\cleardoublepage
\bibliography{bibliography} % Path to your References.bib file
\nocite{*}

\end{document}